\documentclass[fleqn,usenatbib]{mnras}

\usepackage[T1]{fontenc}
\usepackage{ae,aecompl}

\usepackage{graphicx}	
\usepackage{amsmath}	
\usepackage{amssymb}	

\usepackage{txfonts} 

\usepackage{epstopdf}
\usepackage{journals}

\usepackage[none]{hyphenat}

\newcommand{\comm}[1]{}

\defcitealias{2019MNRAS.490.1539R}{R19}

\title[Fractal dimension of optical cirrus]{Fractal dimension of optical cirrus in Stripe82}

\author[A. A. Marchuk et al.]{
Alexander A. Marchuk,$^{1,2}$\thanks{E-mail:a.marchuk@spbu.ru}
Anton~A.~Smirnov,$^{1,2}$
Aleksandr~V.~Mosenkov,$^{1,3}$
Vladimir~B.~Il'in,$^{1,2,4}$
\newauthor
George~A.~Gontcharov,$^{1}$
Sergey~S.~Savchenko,$^{1,2,5}$
Javier~Rom{\'a}n$^{6,7,8}$
\\
$^{1}$Central (Pulkovo) Astronomical Observatory, Russian Academy of Sciences, Pulkovskoye chaussee 65/1, St. Petersburg 196140, Russia\\
$^{2}$Saint Petersburg State University, Universitetskij pr. 28, St. Petersburg 198504, Russia\\
$^{3}$Department of Physics and Astronomy, N283 ESC, Brigham Young University, Provo, UT 84602, USA\\
$^{4}$Saint Petersburg University of Aerospace Instrumentation, Bol. Morskaya ul. 67A, St. Petersburg 190000, Russia\\
$^{5}$Special Astrophysical Observatory, Russian Academy of Sciences, 369167 Nizhnij Arkhyz, Russia\\
$^{6}$Instituto de Astrof{\'i}sica de Andaluc{\'i}a (CSIC), Glorieta de la Astronom{\'i}a, 18008 Granada, Spain\\
$^{7}$Instituto de Astrof{\'i}sica de Canarias, c/ V{\'i}a L{\'a}ctea s/n, E-38205, La Laguna, Tenerife, Spain\\
$^{8}$Departamento de Astrof{\'i}sica, Universidad de La Laguna, E-38206, La Laguna, Tenerife, Spain
}

\date{Accepted XXX. Received YYY; in original form ZZZ}

\pubyear{2020}

\begin{document}
\label{firstpage}
\pagerange{\pageref{firstpage}--\pageref{lastpage}}
\maketitle

\begin{abstract}
The geometric characteristics of dust clouds provide important information on the physical processes that structure such clouds. One of such characteristics is the $2D$ fractal dimension $D$ of a cloud projected onto the sky plane. In previous studies, which were mostly based on infrared (IR) data, the fractal dimension of individual clouds was found to be in a range from 1.1 to 1.7 with a preferred value of 1.2--1.4. In the present work, we use data from Stripe82 of the Sloan Digital Sky Survey to measure the fractal dimension of the cirrus clouds. This is done here for the first time for optical data with significantly better resolution as compared to IR data. To determine the fractal dimension, the perimeter-area method is employed. We also consider IR (IRAS and \textit{Herschel}) counterparts of the corresponding optical fields to compare the results between the optical and IR. We find that the averaged fractal dimension across all clouds in the optical is $\langle D \rangle =1.69^{+0.05}_{-0.05}$ which is significantly larger then the fractal dimension of its IR counterparts $\langle D\rangle=1.38^{+0.07}_{-0.06}$. We examine several reasons for this discrepancy (choice of masking and minimal contour level, image and angular resolution, etc.) and find that for approximately half of our fields the different angular resolution (point spread function) of the optical and IR data can explain the difference between the corresponding fractal dimensions. For the other half of the fields, the fractal dimensions of the IR and visual data remain inconsistent, which can be associated with physical properties of the clouds, but further physical simulations are required to prove it.

\end{abstract}


\section{Introduction}
Cirrus clouds are wispy filamentary structures observed at high latitudes of our Galaxy. They were discovered in the far-infrared (FIR) based on IRAS observations in the early 1980s~\citep{Low_etal1984}. However, much earlier they were also identified at optical~\citep{1955Obs....75..129D,1960Obs....80..106D,1972VA.....14..163D,1976AJ.....81..954S,1979A&A....78..253M,1985A&A...145L...7D,Roman_etal2020} and later at ultraviolet wavelengths~\citep{1995ApJ...443L..33H,Gillmon_Shull2006,2015A&A...579A..29B}. It was also demonstrated that the position of the cirrus clouds is well correlated with the position of some molecular clouds~\citep{Weiland_etal1986,deVries_etal1987,Gillmon_Shull2006}.
\par
The cirrus clouds and molecular clouds, as well as the clouds of neutral hydrogen, are well-known to have a rather complex geometry. They possess hierarchy and self-similarity \citep{Bazell_Desert1988,Dickman_etal1990,Falgarone_etal1991,Hetem_Lepine1993,Vogelaar_Wakker1994,Elmegreen_Falgarone1996,Stutzki_etal1998,Sanchez_etal2005,Sanchez_etal2007,Elia_etal2018,Juvela_etal2018}, that is, they can be described as fractals~\citep{Mandelbrot_1983} or even multifractals~\citep{Chappell_Scalo2001,Elia_etal2018,Beattie_etal2019a,Beattie_etal2019b}. Naturally, an important property of a fractal is its fractal dimension, the value of which characterises how a cloud fills the volume. If it is close to $3$, then the cloud fills the volume like a simple 3D object and if, for example, the cloud consists mainly of linear filaments, the dimension should have a value closer to 1. A number of other approaches to characterise the structure of the clouds was also proposed in the literature, such as the power spectrum analysis~\citep{Stutzki_etal1998,Miville-Deschenes_etal16}, the multi-fractal spectrum analysis~\citep{Chappell_Scalo2001}, the statistical analysis based on the probability distribution function~\citep{Donkov_etal2017} and correlation integral value~\citep{Sanchez_etal2005}. In the present work, we focus mainly on the fractal dimension and do not discuss other approaches further.
\par 
As the fractal dimension characterises the cloud density distribution, its value is determined by the physical processes that govern the cloud evolution. Quite the number of studies is dedicated to exploring the connection between the value of the fractal dimension and the physical parameters of the clouds~\citep{Sanchez_etal2006} including various hydrodynamic~\citep{Federrath_etal2009,Beattie_etal2019a,Beattie_etal2019b} and magneto-hydrodynamic studies~\citep{Kritsuk_etal2007,Kowal_etal2007,Kritsuk_etal2013}. In general, it was found that turbulent flows structure the cloud in such a way that the fractal dimension spans the whole range from $2.0$ to $2.9$ depending on the Mach number value and how the turbulence itself is implemented in simulations~\citep{Kowal_etal2007,Federrath_etal2009, Konstandin_etal2016,Beattie_etal2019a,Beattie_etal2019b}. 
\par
For real clouds observed in astronomical images, a measurement of the fractal dimension is complicated by the fact that the observer usually deals with a 2D intensity distribution with a finite resolution produced by a 3D object with a complicated geometry~\citep{Sanchez_etal2005}. Therefore, strictly speaking, an observational data allows one to only measure the fractal dimension of the projection of a cloud $D_\mathrm{p}$ and not the fractal dimension $D$ of the cloud itself. This problem was thoroughly addressed by~\cite{Sanchez_etal2005} where modelled clouds were considered. An important conclusion of their work is that for a fixed value of the $3D$ fractal dimension $D$, one can obtain different values of $D_\mathrm{p}$ depending on image resolution. \cite{Sanchez_etal2005} also clearly showed that $D$ does not need to be equal to $D_\mathrm{p}+1$, but $D$ tends to be slightly greater than this, e.g. for $D_\mathrm{p}\approx1.35$ it was found that $D$ should be in a range from $2.5$ to $2.7$. It was also found that $D_\mathrm{p}$ should decrease with an increase of $D$ (see their figure 8). In a subsequent work, \citet{Sanchez_etal2007} obtained that some other observational parameters, such as noise level, can also influence the obtained values of the fractal dimension. In recent studies by~\citet{Beattie_etal2019a} and \citet{Beattie_etal2019b}, the authors used hydrodynamic simulations to study how the calculated values of the projection fractal dimension $D_\mathrm{p}$ depend on the volumetric fractal dimension $D$ \textit{and} the Mach number of the flow. It was found that $D$ should be in a range from $D_\mathrm{p}+1/2$ to $D_\mathrm{p}+1$ for high and low Mach number limits, respectively. In contradiction to the results of~\cite{Sanchez_etal2005}, \citet{Beattie_etal2019a} and \citet{Beattie_etal2019b} found that $D$ should increase with an increase of $D_\mathrm{p}$ (see their figure 6). Here we should note these authors calculated the fractal dimension using a mass-length fractal dimension method that differs from the perimeter-area method which is commonly used for measuring the fractal dimension of actual astronomical clouds. Nevertheless, we admit that the simulation setup of~\citet{Beattie_etal2019b} with real clouds and physical processes, which govern their evolution, is much more appealing from a physical point of view than experiments with static geometric fractals with no underlying physics which were considered by~\citet{Sanchez_etal2005}. 
\par
For our convenience, hereinafter we abandon the lower index in $D_\mathrm{p}$ and write it simply as $D$.
\par 
 Most of previous studies on fractal dimension of the cirrus and molecular clouds were carried out based on infrared~IRAS and \textit{Herschel} data~\citep{Bazell_Desert1988,Dickman_etal1990,Vogelaar_Wakker1994,Juvela_etal2018,Beattie_etal2019b}, although a number of other observations were also utilised~\citep{Falgarone_etal1991, Hetem_Lepine1993, Vogelaar_Wakker1994,Sanchez_etal2005}. 
\par
The general approach to measure the fractal dimension, used in those studies, is as follows. First, the observer chooses some contours with a fixed level of intensity (or some set of them), calculates the perimeter $P$ and area $A$ of the structure enclosed by a contour, and approximates the perimeter and area dependence by a power function of the following form: $P \propto A^{D/2}$. This approach turned out to be rather fruitful. In general, it was shown that the structures, measured in such a way, indeed form an almost straight line on the $(\log P, \log A)$ plane with a resultant error on $D$ of about several hundredths and smaller~\citep{Bazell_Desert1988,Dickman_etal1990,Falgarone_etal1991,Vogelaar_Wakker1994,Sanchez_etal2005}. As for the actual values of the fractal dimension, it was found to span a range from 1.1 to 1.7 with preferred values of about $1.3-1.4$ if we account for all results presented in the literature, see Sect.~\ref{sec:comparison}.   
 \par 
 One of the problems with the interpretation of fractal dimension measurements is that different data can have different image resolution (pixel scales), as well as some other observation specifications, such as different point spread functions (PSFs), etc. As was already mentioned, \citet{Sanchez_etal2005} showed that differences in resolution can affect the obtained values of the fractal dimension and, in some cases, a low resolution can even lead to a significant underestimation of the fractal dimension (see figure~7 in~\citealt{Sanchez_etal2005}). In terms of this problem, it is important to examine how the previous results hold if we use some new data with better observational specifications.
 \par
 In recent work by~\cite{Roman_etal2020}, the authors distinguished a number of cirrus clouds in several fields of the Sloan Digital Sky Survey (SDSS) Stripe82 \citep{Abazajian_etal2009,Fliri_Trukillo2016} after some additional image processing, which included creating mosaics, masking, and subtracting the sky background and instrumental scattered light. In the present article, we use the same fields as presented in~\cite{Roman_etal2020}, where the cirrus clouds have already been distinguished, to calculate their fractal dimension. In this we pursue two goals. 
 \par 
 The first goal is to actually calculate the fractal dimension of the clouds in the optical. To our knowledge, measuring the fractal dimension in the visible has not been done before. At the same time, most of previous investigations used IR data as a source material with much worse resolution. As mentioned above, these factors are important for fractal dimension measurements. We  should also note that, from a physical point of view, it is important to explore whether the results of fractal dimension measurements depend on wavelength. And if they do, that means that the upcoming physical simulations of the clouds should take this into account when comparing the simulated data with actual observational data. In this study, we also compare the obtained results for the optical with those obtained for the IR for the same fields and, in general, with the results of previous works.  
 \par 
 The second goal of this paper is rather methodological. Optical images have better resolution and, hence, there are many small-scale features present on such images that cannot be detected in the IR. Thus, optical data can be used to directly measure to what degree the various effects, such as image and angular resolution (PSF),  can change the fractal dimension values. This is also important for future studies of the geometric properties of dust clouds. 
 \par 
 On the other hand, although the use of optical images has many advantages, bright sources in the optical are much more numerous as compared to the IR. Masking such sources often introduces empty areas which can shred up the clouds one tries to measure. For this reason, to reliably measure the fractal dimension, one should study how the masking itself affects the measurement. We pay special attention to this matter in course of the present article and prepare several additional simple experiments to estimate the effect of masking. 
\par 
The structure of the present paper is as follows.
In Section~\ref{sec:data}, we describe the optical and IR data we used to calculate the fractal dimension of the clouds.
In Section~\ref{sec:methods}, we provide a thorough description of our algorithm to calculate the fractal dimension and give a description of a Monte-Carlo simulation setup which we used to estimate the errors for our fractal dimension measurement, as well as to estimate how various observational specifications affect the results.
In Section~\ref{sec:results}, we present results of our fractal dimension measurements for the optical and IR data and compare the obtained results with those from the literature.
In Section~\ref{sec:effects_that_change_D}, we discuss the influence of various effects on our fractal dimension measurements, including angular resolution and masking.
In Section~\ref{sec:discussion}, we discuss the physical reasons for the observed differences of the fractal dimensions in the optical and IR and compare our results with those obtained in previous studies. 
We summarise our results in Section~\ref{sec:sum}.

\section{Data}
\label{sec:data}

\begin{figure}
\includegraphics[width=0.95\columnwidth]{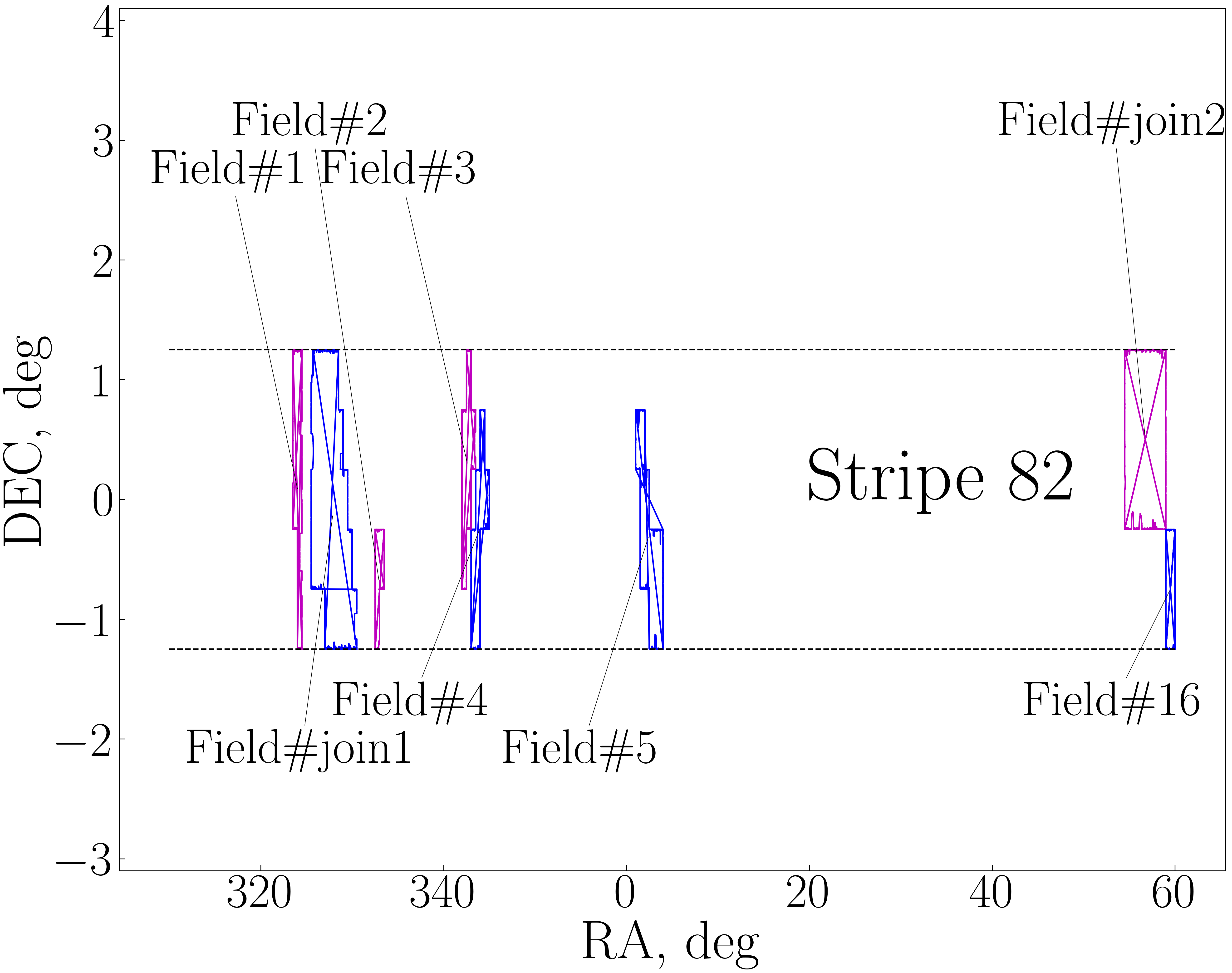}
\caption{Position of the fields under consideration in the plane of the sky. The dashed lines depict the Stripe82 borders. The frames of the fields are colour-coded for a better distinction of adjacent fields.}
\label{fig:skypos}
\end{figure}

\subsection{Optical data}
For optical data, we use 16 different fields in the $g,\,r,\,i,$ and $z$ bands which were prepared by~\cite{Roman_etal2020} to specifically distinguish and study cirrus clouds. A detailed description of each field is summarised in table~2 of \cite{Roman_etal2020}. Here we briefly provide some important details on the data reduction pipeline, which was designed to carefully probe cirrus clouds in the Stripe82 fields, and on our further processing. 
\par
The original raw fields were taken from the SDSS Stripe82 database~\citep{Abazajian_etal2009}. The Stipe82 data itself is obtained with the 2.5-meter telescope of the Apache Point Observatory and covers a stripe in the sky with an angular area of 275 square degrees within $-50^\circ < \mathrm{RA} < 60^\circ$ and $-1.25^\circ<\mathrm{Dec.}<1.25^\circ$. The exposure time was about one hour for each field. The original pixel size of the images was 0.396 arcsec per pixel.
%
In addition to the standard SDSS data reduction, the fields were stacked by~\cite{Fliri_Trukillo2016}, who tried to carefully preserve the characteristics of the background which represents a sum of several diffuse light components. Then, the residuals of the co-adding process were removed in~\cite{Roman_Trujillo2018}. In~\cite{Roman_etal2020}, the fields were further processed to account for the instrumental scattered light produced by the extended PSF wings of the stars. Also, in~\cite{Roman_etal2020}, accurate masking of all sources that contaminate the cirrus emission was carried out and the fields were rebinned from the original SDSS pixel size to 6 arcsec per pixel. The relative location of all fields in the sky is shown in Fig.~\ref{fig:skypos}. As can be seen from this figure, the fields are distributed all over Stripe82 and do not lie in one area or direction. The fields span the galactic latitude $b$ between $-35^\circ$ and $-61^\circ$ for Field\#1 and \#5, respectively.

\par

One important difference from~\cite{Roman_etal2020} is that we have 8 large fields instead of their smaller 16. The reason is that we decided to compose the overlapping fields into a single mosaic because the total image area can be an important factor for measuring the fractal dimension of a cloud. Specifically, we merged two sets of the fields, namely Fields\#6-10 and \#11-15. These fields form a sequence of tiles that are located next to each other in the original Stripe82 data. We refer to the resulted composed fields as `\#join1' and `\#join2' in our further discussion. The notation of the remaining Fields\#1-5 and \#16 stays unchanged. The reason why they were cut in the first place is that \cite{Roman_etal2020} aimed to have more points with different densities of dust to better sample the correlation between density and optical colours. 

There is one major problem with the large fields which stems from the SDSS data reduction pipeline: it can cause over-subtraction of the diffuse light on a scale of several acrminutes, thus not allowing to correctly investigate cirrus clouds of similar sizes. More about this issue can be found in the original paper by~\cite{Roman_etal2020} and later in Sect.~\ref{sec:discussion}. 

\subsection{IR data}
\begin{figure*}
\includegraphics[width=1.95\columnwidth]{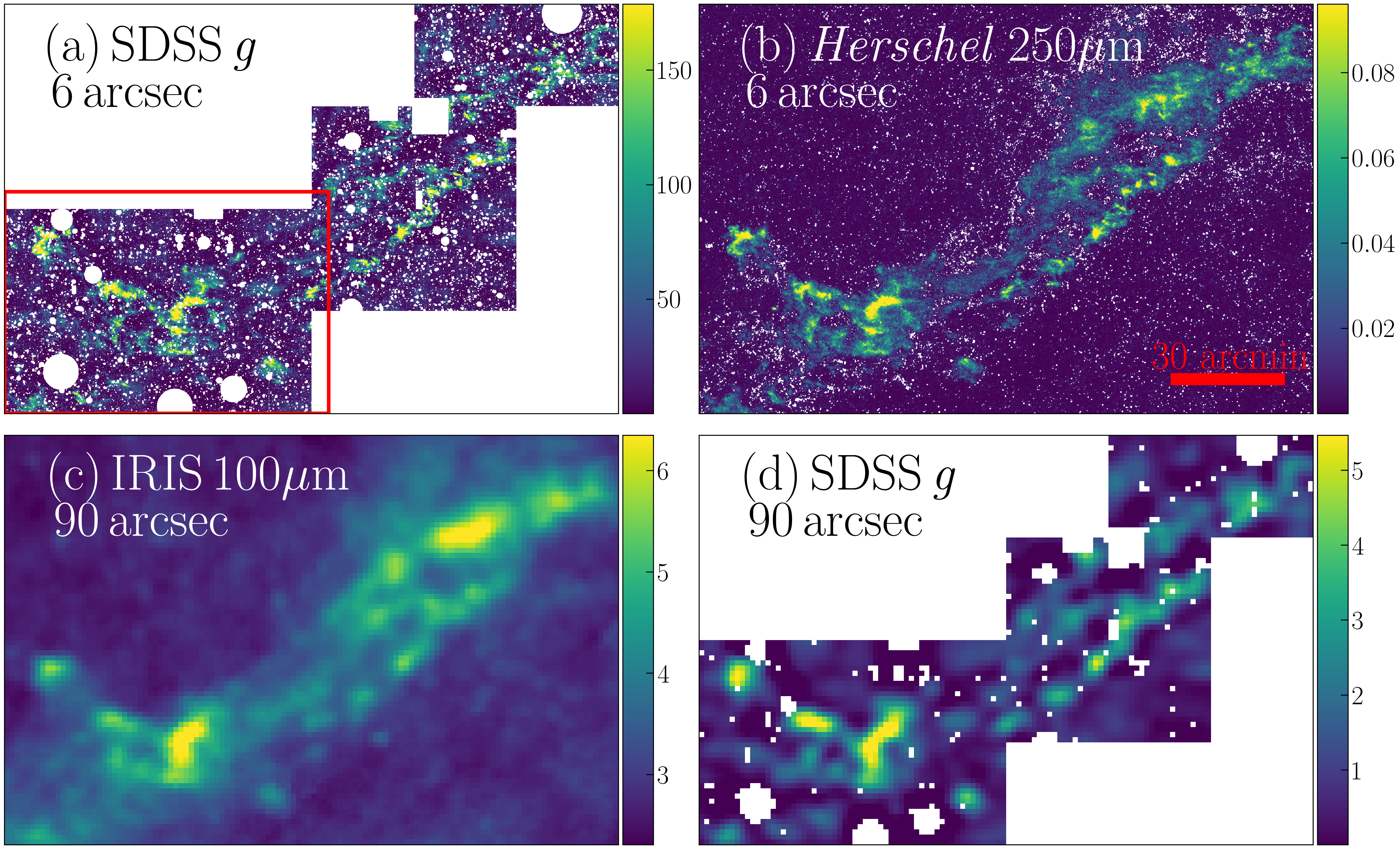}
\caption{IR and optical data counterparts for Field\#5. All images show the same region. Panel (a) displays the data in the $g$ band with the 6~arcsec resolution and with the colourbar in counts. The red frame limits the part which is later shown in Fig.~\ref{fig:pipe}. Panel (b) showcases the \textit{Herschel} counterpart with the same resolution but with a different mask (the colourbar units are MJy/beam). Panel (c) represents the IRIS 100~$\mu$m data with the 1.5~arcmin resolution and in units MJy/sr. Finally, panel (d) shows the data in the $g$ band, rebinned to the 1.5~arcmin pixel size and convolved with the IRIS PSF, the units are counts. Note that the IRIS data in panel (c) is depicted without mask, which is exactly the same as the one in panel (d).}
\label{fig:f5_comp}
\end{figure*}

Since most of previous studies on the fractal dimension utilised IR data, it is interesting to compare the fractal dimension of the clouds that we observe in the optical with their counterparts in the IR. To this aim, we use the Improved
Reprocessing of the IRAS Survey (IRIS, \citealt{IRIS}) and \textit{Herschel} \citep{Viero_etal2014} data. IRIS provides reprocessed IRAS data with a slightly improved angular resolution (4.3 arscmin at 100~$\mu$m), better calibration and zodiacal light subtraction.  For all of the optical fields we extracted their IR counterparts from the IRIS 100~$\mu$m database using special {\small IDL} routines, which were designed to produce IRIS  mosaics\footnote{Available at \url{https://www.cita.utoronto.ca/~mamd/IRIS/IrisDownload.html}}. For Field\#5, we also analyse the~\textit{Herschel} 250~$\mu$m data from~\cite{Roman_etal2020} where the \textit{Herschel} Stripe82 Survey (HerS, \citealt{Viero_etal2014}) was used to identify whether the diffuse emission, observed in the optical, is due to the Galactic dust or there are some other sources responsible for it. This field has its own mask to filter out all sources, which are not related to dust, see fig.~7 in \cite{Roman_etal2020} and panel (b) of Fig.~\ref{fig:f5_comp}. This figure also shows an optical image in the $g$ band and the IRIS data, so they can be directly compared with each other. Note a good congruence of the cirrus contours in the separate subplots and a difference between the masked regions in panels (a) and (b).
\par 
To ensure that our processing procedure is not affected by some internal flaws, we also exploit some IR fields for which $D$ has been measured in previous studies. We consider two of the five IRAS 100$\mu$m fields studied in~\cite{Dickman_etal1990}, namely the Chameleon and Taurus fields. We select an area within RA and Dec. coordinates as in \cite{Dickman_etal1990}. The reason why we consider only these two specific fields from~\cite{Dickman_etal1990} is because they are located inside a single IRAS plate and, thus, there is no need to compose several areas to produce a resulted image for our comparison analysis. 
\par 
There are several major points we should mention concerning the IR data. The IRIS data has a relatively low angular resolution with a pixel size of 90~arcsec. The IRIS PSF is much more extended than the PSF for the optical data, with the characteristic scale FWHM = 4.3 arcmins. As we show below, both these factors are important for measuring the fractal dimension. The \textit{Herschel} data has a much higher image resolution (a pixel size of 6 arcsec) and a much better PSF (FWHM = 18.1 arcsec). 
\par
Below we consider two types of IR images: one with the optical mask applied and the other without any mask. The latter option should be considered to correctly compare the results of the present study with the results from the literature. In most previous studies a masking procedure was not actually carried out (see~\citealt{Bazell_Desert1988,Dickman_etal1990,Juvela_etal2018}). Instead, all the external sources were usually cut out from the analysis based on their size, that is, all contours, which were smaller than some fixed size, were excluded after the perimeters and the area of the clouds had been measured. 

\section{Methods}
\label{sec:methods}
\subsection{Method description}

In order to compute the fractal dimension $D$, we employ a perimeter-area method following \cite{Bazell_Desert1988,Dickman_etal1990,Falgarone_etal1991,Hetem_Lepine1993,Vogelaar_Wakker1994,Sanchez_etal2005,Juvela_etal2018}. To measure the fractal dimension of a cloud, one should find the power index in the relation between the cloud perimeter $P$ and its surface area $A$, which has the following form \begin{equation}\label{eq:AP} 
P =K\times A^{D/2}\,,
\end{equation}
where $D$ is the sought value of the fractal dimension and $K$ is the intercept coefficient. While the fractal dimension $D$ can be directly connected to some physical properties of the cloud (see Introduction), the intercept $K$ should characterise the general shape of the cloud. However, we note that, based on the results of previous studies~\citep{Dickman_etal1990,Vogelaar_Wakker1994}, the intercept values obtained for the real clouds can hardly be used to associate the shape of the molecular clouds with the shape of some simple geometric objects like ellipses or circles (see discussion section in~\citealt{Dickman_etal1990}).
\par
The above described method cannot be applied as is because the optical fields have a significant number of masked sources within individual clouds. In principle, this circumstance can change the actual fractal dimension in some non-trivial way. For example, suppose we have some data which satisfies eq.~(\ref{eq:AP}). If we start extracting ``holes'' from such clouds, then the resultant dependence will have a greater fractal dimension $D=2\log P/\log A$, because their area will decrease, while the perimeter will increase. Thus, some additional steps are needed to mitigate this effect.

\begin{figure*}
\includegraphics[width=1.95\columnwidth]{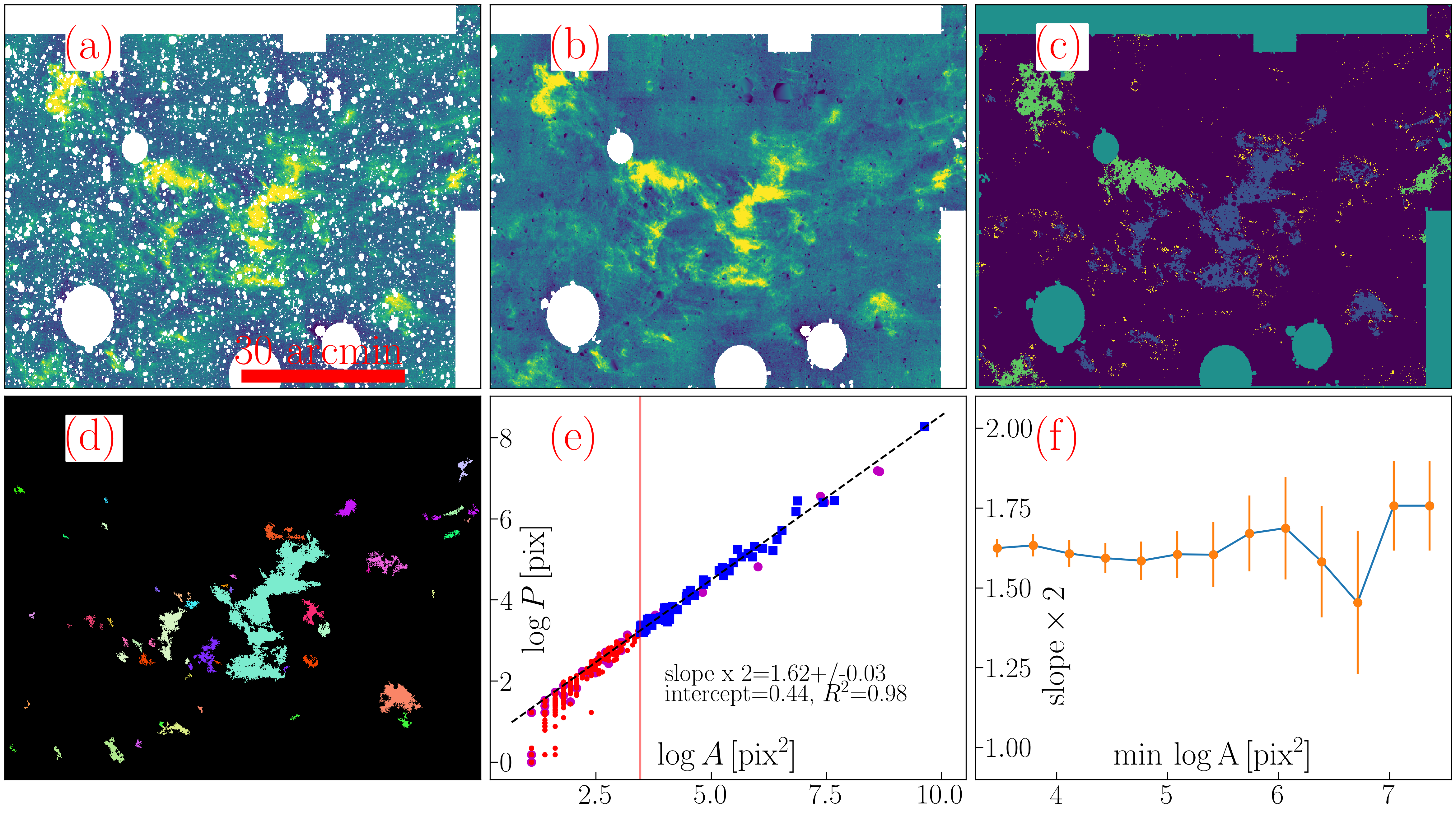}
\caption{An illustration of the method pipeline for part of Field\#5 shown in Fig.~\ref{fig:f5_comp} and for the selected contour level=$27$~mag arcsec$^{-2}$. 
For the description of the individual panels see Sect.~3. In panel (c) blue colours show the selected contours for a further analysis, light green -- contours, which touch the mask, deep green -- large mask and image borders, yellow -- small filtered clouds with an area $A<30$~square pix. Different clouds in panel (d) are shown by different colours, the ``holes'' are filled (see text for details). In panel (e), blue squares correspond to individual clouds from (d), magenta circles are the clouds which touch the large mask (see step (vii) for details) and then filtered out, red points -- all small segments, which are filtered out by the `tailcut' parameter, depicted by the solid vertical line. Note that points from only part of the original Field\#5 are shown and, thus, the derived slope is just an example and not the real $D$ for the whole field.}
\label{fig:pipe}
\end{figure*}

The sequence of steps for $D$ estimation for each image is illustrated in Fig.~\ref{fig:pipe} and described below in the following nine-points list. Hereinafter, values of $A$ and $P$ are given in square pixel and pixel units, respectively, unless otherwise explicitly stated. Throughout the text all logarithms are natural. It is important to note here that some steps are optional, while other depend on subjective parameters. In the next subsection, we take them into account using Monte-Carlo simulations to verify whether the choice of the parameters can affect a fractal dimension measurement. The following list summarises the details of our algorithm:

(i) We select extended parts of the mask which are, at first sight, too large to be properly interpolated inside. These parts appear as  large white areas in panel (b) of Fig.~\ref{fig:pipe} (also see panel (a) and (d) in Fig.~\ref{fig:f5_comp}). The second reason, why this step is necessary, is that such masked areas can, along with the image borders, shred up individual cirrus clouds or cut off some of their parts. If this happens, the $D$ value can change.

(ii) This step is optional and includes an interpolation of all mask parts which have not been previously selected in step (i). We use a linear interpolation method which produces an overall smooth image as shown in panel (b) in Fig.~\ref{fig:pipe}. Note a small number of artefacts for the ``holes'' of medium size.

(iii) Then we select a brightness contour level which corresponds to a lower boundary of the cirrus emission. All pixels of individual clouds, which we consider in the next steps, should have an equal or greater brightness. All the selected contours for the level $27$~mag arcsec$^{-2}$ in the $r$ band are displayed in panel (c) of Fig.~\ref{fig:pipe} by blue, light green and yellow colours and the difference between them is explained below.

(iv) This step is optional. We test the findings of \cite{Roman_etal2020} that cirrus can be filtered out from other sources by its optical colours. Thus, we apply an additional mask to select only those pixels that satisfy the $(r - i) < 0.43 \times (g - r) - 0.06$ condition (see equation (1) in \citealt{Roman_etal2020}). We note, however, that all emission regions from the analysed fields are expected to be actual cirrus clouds, thus these measurements are merely performed out of methodological interest, and the results obtained this way are not incorporated in the final $D$ estimation. We discuss this matter more thoroughly in Sec.~\ref{sec:color}.

(v) As mentioned above, additional ``holes'' inside cirrus clouds due to a mask can increase the fractal dimension. In this step, we decide to ``fill in'' all mask parts which are completely surrounded by the pixels that have been previously found to belong to some clouds. Hence, we assume that such mask parts are actual parts of the cirrus. If the ``holes'' have already been interpolated, then this step has no effect regardless of the choice to ``fill in'' or not. We note that individual clouds can still have ``empty holes'' inside their body. Such holes are not due to the masking of foreground sources. They appear as a result of the combined effect of the cloud geometry, projection effects, and the selected contour level (see figure~1 in \citealt{Sanchez_etal2007}). Panel (d) of Fig.~\ref{fig:pipe} showcases how the clouds, distinguished in panel (c) of the same figure, are transformed after applying the described procedure. 

(vi) All the selected pixels are segmented to individual clouds using simple connectivity rules, where two pixels can be connected by a common side, but not by a common corner. Individual clouds are represented by different colours in panel (d). One can note the variety of shapes and sizes of individual segments.

(vii) The extracted individual clouds are then filtered in two different ways. First, we remove all noisy segments, the area $A$ of which is less than some small value, e.g. 5 square pixels (panel (e) of Fig.~\ref{fig:pipe}, shown in yellow).  Second, we choose whether we should use all the segments which touch a border of the image or the extended mask parts selected in step (i) even by one pixel. This is explained by the fact that one can only analyse part of such a cut-off cloud. If we consider only part of a cloud, its fractal dimension $D_\mathrm{part}$ can differ from the fractal dimension $D$ of the whole cloud. This effect is investigated in greater detail in Sect.~\ref{sec:discussion}. The segments, removed after the first and second procedures, are shown in yellow and light green in panel (c) of Fig.~\ref{fig:pipe}, respectively.

(viii) For all remaining clouds, we find the area $A$ and perimeter $P$ using the method {\small regionprops} from the {\small skimage} package. The linear regression is fitted to data ($\log A$, $\log P$) using an ordinary least-squares method. The slope of the regression $D/2$ provides us with the desired fractal dimension value, as well as the intercept $K$. This is illustrated in panel (e) of Fig.~\ref{fig:pipe}. We also depict the points for small (red) and sliced (magenta) clouds, filtered out in the previous step. We note that the cut-off clouds follow the regression line fairly well. In contrast, the inclusion of ``small'' noisy clouds into the fit bends down the lower left tail of the regression line, which leads to an increase of the $D$ value. This is due to a much lower perimeter-to-area ratio of such noisy clouds with respect to larger ones.

(ix) Since the decision of which clouds should be filtered out because of their size is, strictly speaking, an arbitrary one, we introduce an additional parameter which we call the `tailcut' parameter. This parameter defines which portion of the small clouds in the regression should be removed from our analysis. The value of the parameter can be set in a range from 0 to 1 assuming that 0 corresponds to the lower limit of the $ \log A $ range, and 1 corresponds to the upper one. The `tailcut' parameter has a simple geometric interpretation. In panel (e) of Fig.~\ref{fig:pipe}, its value can be represented by a vertical line. By moving this line to the right, i.e. increasing the `tailcut', one can measure how stable the regression fit to this sort of filtering is. The example of this process is shown in panel (f) of the same figure. The error bars correspond to the errors of the linear regression fitting. We note that, for this selected part of the cirrus, the measurement of the fractal dimension is stable within the margin of error until we start filtering out clouds with an area $\log A > 5.5$, or $A > 250$ square pixels.

For the IR counterpart images from the \textit{Herschel} and IRIS surveys, step (iv) was excluded from the sequence, since we would like to use the IR data as is for better comparison with the literature. 
Note, that, contrary to the previous studies, we decided not to use all the levels for a given field image to fit one single regression line. Instead, we measure the fractal dimension for each brightness level under consideration. This produces a less stable measurement, but, at the same time, gives us a better understanding of whether the fractal clouds have the same properties at different brightness levels in the optical bands. 
\par
The presented method can be used as is. However, to estimate the effect of individual steps on a fractal dimension measurement and to estimate its uncertainties, we run a Monte Carlo (MC) simulation, the setup of which is described below.

\subsection{MC simulation}
\par
\begin{figure*}
\includegraphics[width=1.95\columnwidth]{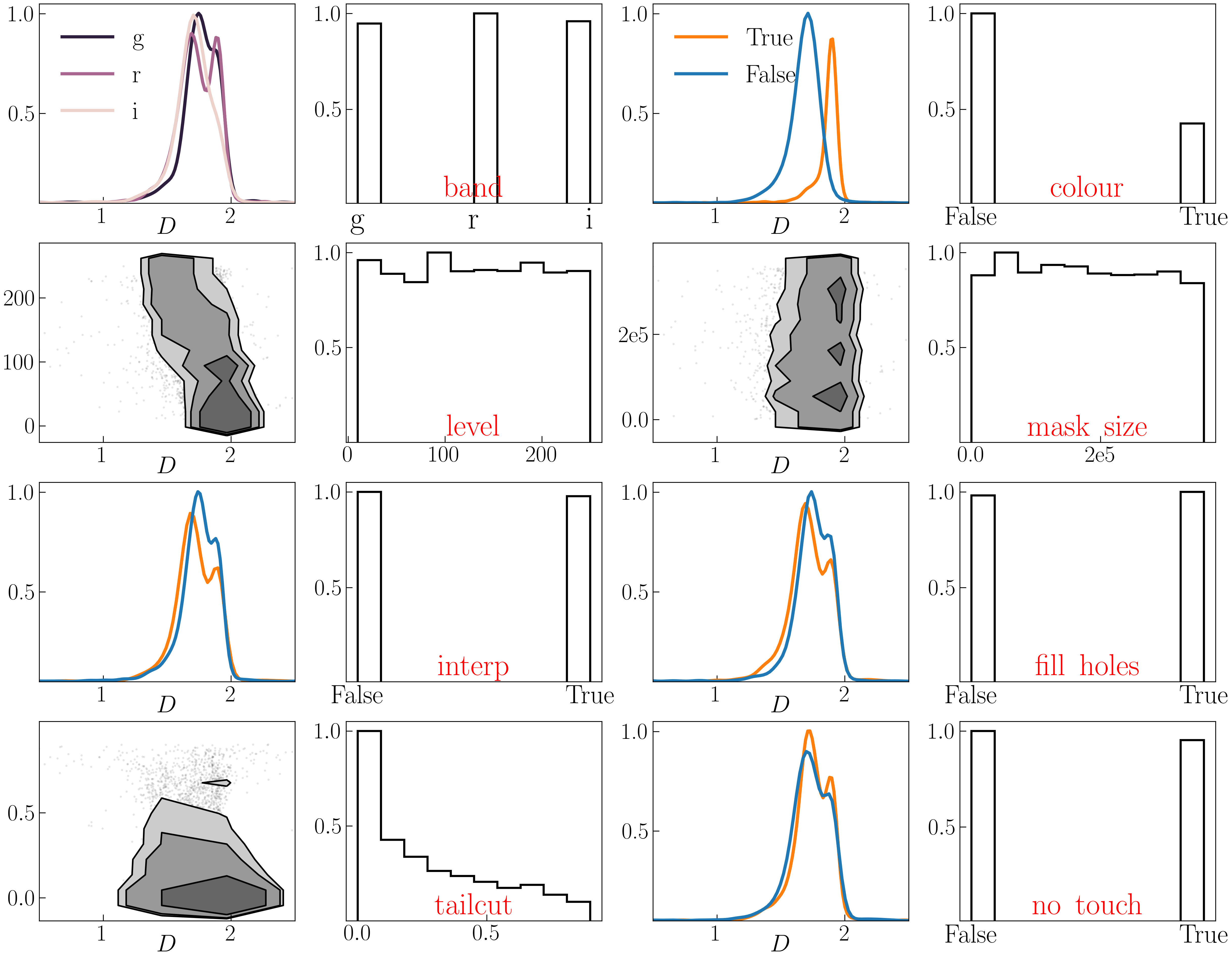}
\caption{Dependence of the fractal dimension $D$ on the MC parameters for Field\#3. Each parameter is illustrated by two subplots. The first subplot shows how parameter values affect the resulted $D$ estimation, while the second subplot represents the parameter distribution using the corresponding histogram. For discrete parameters, the data is presented as a $D$ density distribution for each option; for continuous parameters --- as standard deviation density contours, where individual dots are outliers. All histograms are normalised in such a way that the height of the largest bar equals unity and all density contours are drawn for the same bins as in the related histogram but slightly smoothed. For continuous parameters, the vertical limits in the left plot are the same as the horizontal ones in the right plot. In density plots, the blue line is for {\it False} option and the orange one is for {\it True}. The parameters have the following notation: `band' is for data in an optical band (the $z$ band is excluded, see Sec.~\ref{sec:diffbands} for details), `colour' is for colour filtering, `level' is for brightness contour in counts, `mask size' is for large extended mask size limits, `interp' is for option whether interpolate mask or not, `fill holes' parameter regulates whether we should fill the mask inside the contours or not, `tailcut' is the same as in the main text, `no touch' is to filter out clouds which are likely to be shaded by the extended mask, see step (vii).}
\label{fig:mcparamdep}
\end{figure*}

For each optical field, we generate a random set of parameters, according to the rules described below, and measure the fractal dimension $D$, intercept $K$, and their errors. The regression line cannot be fitted for some combinations of the parameters, e.g. when all clouds are filtered out. We consider such realisations unsuccessful and do not take them into account. We run 10000 successful realisations and estimate an average value of $D$ and its $1\sigma$ error. In Fig.~\ref{fig:mcparamdep}, one can see a typical example of how individual steps adjust the $D$ value for Field\#3. A detailed description for each of these steps is as follows.

First, we randomly choose which optical band from $g,r,i,z$ will be used in this particular realisation. In this step, we assume that all bands should equally contribute to the final estimation of the fractal dimension. However, in the course of this study, we found that the results, obtained for the $z$ band, significantly differ from others. This can be interpreted by the fact that the $z$ band is much shallower as compared to the $gri$ bands: the surface brightness limit $\mu_{lim}(3\sigma; 10\arcsec\times10\arcsec)=29.1,28.6,28.2,26.6$ mag\,arcsec$^{-2}$ for the $g$, $r$, $i$ and $z$ bands, respectively (see \citealt{Roman_etal2020}). Therefore, in our further discussion we do not take into account the results obtained for the $z$ band. This matter will be addressed in greater detail in Sec.~\ref{sec:diffbands}. Second, we choose with equal probability whether we should include the cut-off clouds or not. Next, the size of the extended mask parts, described in step (i), is randomly picked out from an interval of 200 square pixels to one third of the field area. Each mask segment, the area of which is larger than the selected size, is then considered to be too large to be interpolated. 
We add a thin frame with a width of 5 pixels around each field. This frame is also used as an additional mask to filter out clouds in step (vii) if they touch it. Next, we select whether we should interpolate all the remaining mask parts or not (see step (ii)) with the equal probability.  As regards the interpolation method, we found that the choice of the method used has a negligible or no effect on our results and, thus, we can use a linear interpolation for simplicity. 

The brightness contour level is selected from a uniform distribution within a range from 0 to the maximum value. Since we work with images in different bands, the distribution range is not the same for all bands, with a trend of the maximum brightness to increase towards the $z$ band. We decided to use a wider range of brightnesses, which correspond to all individual bands and all possible fields (even if it is possible that some image has 0 pixels for a selected brightness level, because this situation is rare and we do not take it into account in the final measurement anyway). It is important to note that this is not true for the IRIS counterparts, where some fields have a zero intersection in the brightness levels. Next, we decide whether to filter out a cirrus or not to do that based on its colour with the probability $p=0.66$\footnote{We have shifted the probability distribution here because the colour filtering  is an experimental step which has not been done before. Thus, we decided to collect more realisations without such filtering.} for the latter (see step iv). Also, we decide whether to fill in the inner mask ``holes'', as explained in step (v). The `tailcut' parameter from step (ix) is picked out randomly from the $[0.0, 0.95]$ range and then squared, in order to shift the distribution to filter out small clouds more frequently than larger ones. We note that we do not choose the size for the initial filtering of small clouds in step (vii) at random, but conservatively filter out all segments with $A \leq 5$~square pix to speed up the process. The reason behind this is that the `tailcut' parameter does the same filtering. Finally, we do not perform a regression fit if there are three or less points to fit. Such realisations are considered unsuccessful.

As was already mentioned in the previous section, for the IRIS and \textit{Herschel} data, the colour filtering is not applicable and, hence, it is not carried out in our MC simulation too. Another important difference from the optical data is that because of the worse resolution, only a constrained set of the parameters for the IRIS fields can lead to a successful measurement of the fractal dimension and, therefore, it is difficult to collect all 10000 realisations. Moreover, since some fields are extensively brighter and some images contain only a small number of clouds, selecting cirrus contours from the same distribution often results in an empty sample. Thus, for $100~\mu$m counterparts we decided to run the same number of realisations as for the optical bands, neglecting the fact whether a particular realisation is successful or not.  All other parameters in the MC simulation for the IR counterparts remain the same.


\section{Results}
\label{sec:results}

\begin{figure*}
\includegraphics[width=1.95\columnwidth]{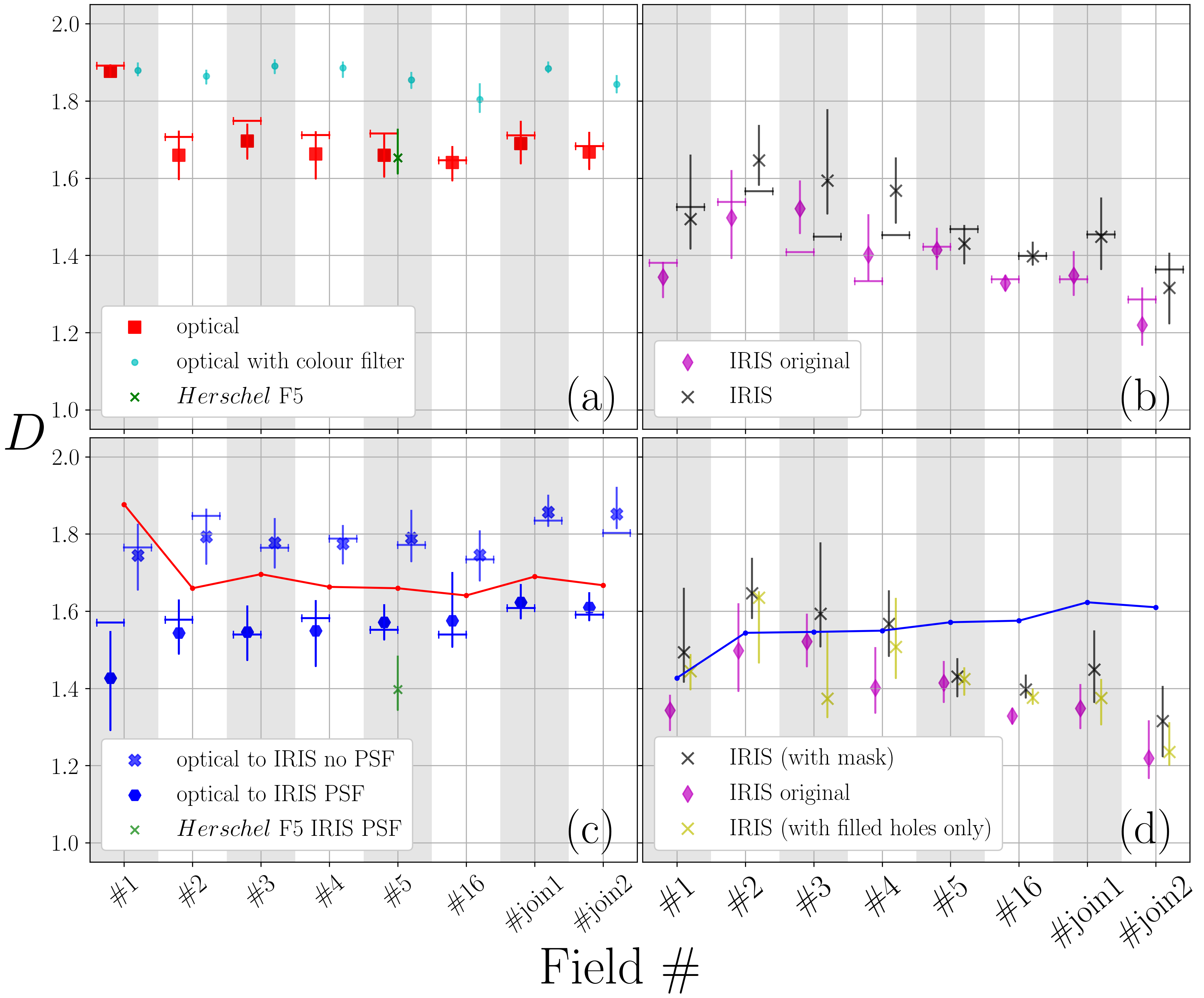}
\caption{Fractal dimension values obtained from our MC results for the different data. For each field the points show a median value and the vertical lines correspond to a 1$\sigma$ error. The horizontal segments show the result when the contours from all MC realisations are placed on the regression line (see text for details). Panel (a) shows all data with the 6~arcsec resolution, i.e. the optical data is shown in red, the optical data with colour filtering is depicted in cyan, and green is for \textit{Herschel} Field\#5 250~$\mu$m (see discussion in Sec.~\ref{sec:color}). Panel (b) contains the results for the IRIS counterparts (here, the magenta and grey colours depict the original data and the data with the optical mask, respectively). Panel (c) illustrates how $D$ changes for the optical (light blue) fields after rebinning to 90~arcsec and after applying the IRIS PSF (deep blue), the green point is for the single \textit{Herschel} field (see discussion in Sec.~\ref{sec:psf} for both (b) and (c) panels). Finally, panel (d) is the same as (b), except that it also depicts the effect of partial mask filling, which is shown using yellow-coded data points (see discussion in Sec.~\ref{sec:masking}). The red line in panel (c) corresponds to the optical data from (a), the blue line in panel (d) corresponds to the optical data with the IRIS PSF in (c).}
\label{fig:allmc3}
\end{figure*}

Table~\ref{table:main_parameters} and Fig.~\ref{fig:allmc3} summarise the results of our fractal dimension measurements from the MC simulation for the clouds from the selected fields. 
We note that these table and figure present not only results of direct fractal measurements of optical data, but also  results of other measurements that are important for the current analysis. Namely, the columns 5 and 6 present the fractal dimensions and intercepts for the optical and \textit{Herschel} data with the image resolution reduced to the IRAS data (panel (c) of Fig.~\ref{fig:allmc3}), while the columns 7--10 show the values of fractal dimension $D$ and intercepts $K$ for the IRIS counterpart with the optical or \textit{Herschel} mask applied where possible (columns 7 and 8, panel (b) of Fig.~\ref{fig:allmc3}, black crosses) and without the mask (columns 9 and 10, panel (b) of Fig.~\ref{fig:allmc3} magenta diamonds). The values presented in the columns 5--10 are provided  to validate our approach and also to make comparison with previous studies more direct. In this section, we mainly discuss the new results of fractal dimension measurement that are done based on the optical data only (columns 3 and 4, panel (a) of Fig.~\ref{fig:allmc3}), while we postpone a detailed description of the other measurements and how they were carried out to the following section, where we analyse the various factors that contribute to a fractal dimension measurement.
\par
For optical data, the fractal dimension value, averaged across all fields, is $\langle D \rangle=1.69^{+0.05}_{-0.05}$ with a remarkably small characteristic spread of values $\sigma(D)=0.02$, if the outlier for Field\#1 is not taken into account ($\sigma(D)=0.07$ if it is taken into consideration). For the same clouds, measured in the IRIS fields, the fractal dimension is considerably smaller $\langle D \rangle=1.48^{+0.10}_{-0.07}$, $\sigma(D)=0.10$ with the optical mask applied and $\langle D \rangle=1.38^{+0.07}_{-0.06}$, $\sigma(D)=0.09$ without it. A typical measurement error obtained for all MC simulation realisations appeared to be smaller than 0.1 for both the optical and IR fields. Note, that the reported average values are calculated as a mean of the $D$ values from Table~\ref{tab:fd_result} without any correction for the individual size of a particular field or the number of clouds.
\par
\begin{figure*}
\includegraphics[width=1.95\columnwidth]{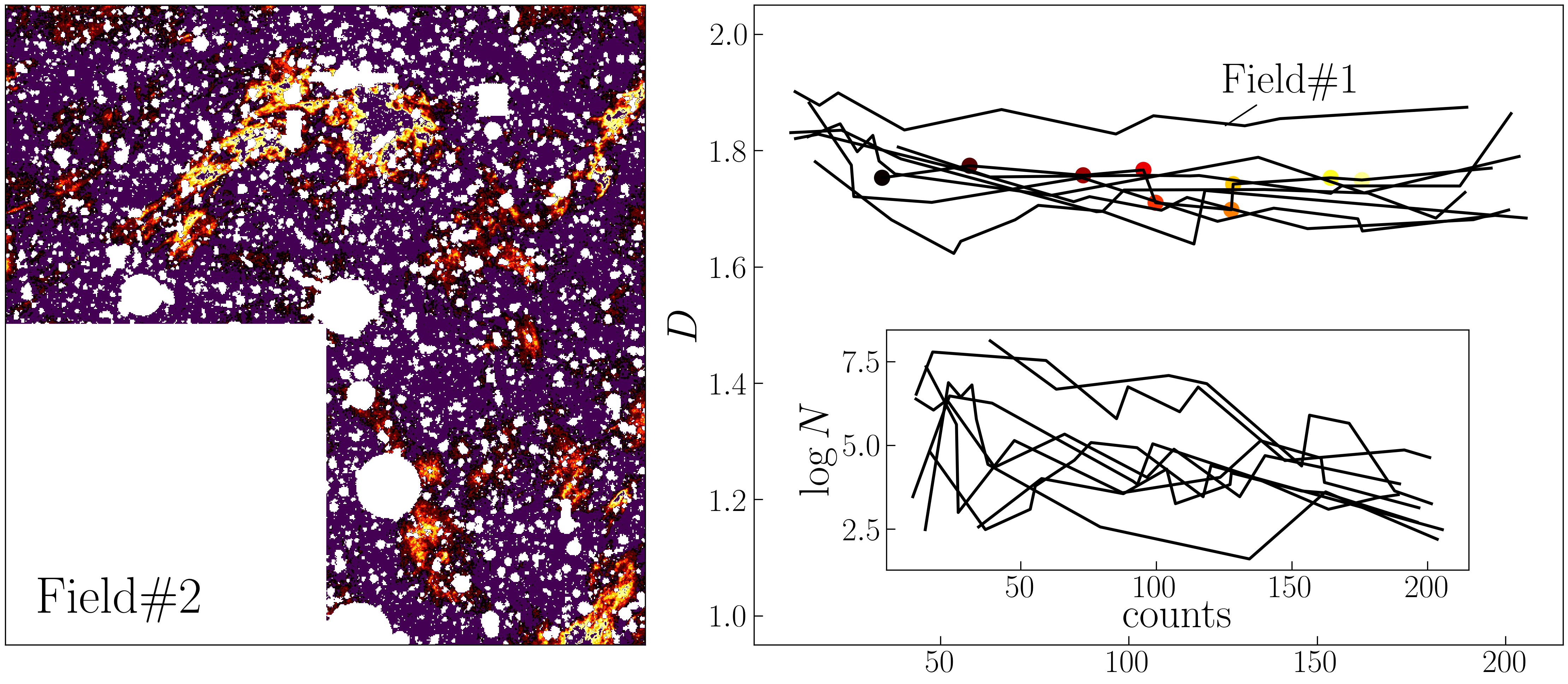}
\caption{Illustration of how the fractal dimension $D$ remains constant within the selected contour level change. The left subplot shows Field 2 as an example of different contours, where a brighter colour relates to a higher brightness. White pixels represent the mask used. The right panel shows how $D$ depends on the contour level, each line corresponds to an individual field, coloured dots show the results for Field\#2 and the same levels as on the left. A smaller subplot on the right side shows the logarithm of the number of clouds for the corresponding contour level.}
\label{fig:lev_dep}
\end{figure*}

 
First of all, we should emphasise that the measured fractal dimensions appear to be less than 2 for all the cirrus clouds under consideration. At the same time, this value does not change much if we consider different spatial scales. This claim is illustrated in panel (e) of Fig.~\ref{fig:pipe} for Field\#5, where it can be seen that the dependence of the perimeter on the area is clearly has small scatter from the best-fit line, and in Fig.~\ref{fig:lev_dep} for all fields, which illustrates that $D$ remains almost constant while the brightness of the contours and the scale vary. Therefore, we obtain that all the selected cirrus clouds are of a fractal nature. While this result is not conceptually new and has been known for a long time starting from~\cite{Bazell_Desert1988}, it is still important to note because, strictly speaking, 1) we analyse the clouds which have not been considered for measuring their fractal properties, and, most importantly, 2)  we mainly concentrate on optical cirrus, whereas most of the previous studies were focused on IR data.
\par 
Our value $\langle D \rangle=1.38^{+0.07}_{-0.06}$, averaged over all IRIS fields with no mask, seem to be slightly greater as compared to the average values obtained in previous works. For convenience, in Table~\ref{tab:previous} we collected all the results of fractal dimension measurements from previous studies where the same perimeter-area based method was used. As one can see, depending on the data, there is a range of values the fractal dimension can take. For example, \cite{Juvela_etal2018} obtained that the average fractal dimension of the dust clouds is about $D\approx1.2$ and spans a range from $1.05$ to $1.40$ based on the \textit{Herschel} data. \cite{Bazell_Desert1988} measured the fractal dimension specifically for cirrus clouds using IRAS data and found that the clouds have a fractal dimension of about $D\approx1.26$, with a range from $1.12$ to $1.4$. \citet{Vogelaar_Wakker1994} found for IRAS data that the fractal dimension of different clouds ranges from 1.2 to 1.6. While our average $\langle D \rangle$ seems to be slightly larger than those, the values, obtained for individual clouds, seem to fall into the same range of values. Two possible exceptions are Field\#2 and \#3 where the fractal dimension $D$ is about 1.5, which is around an upper limit of the values obtained in the literature, but, overall, our values of $D$ are still consistent with those from~\cite{Vogelaar_Wakker1994}.
\par
For the same IR data but with the mask applied, the $D$ values become greater by about $0.1$, but even in such a case they are still within an interval from 1.2 to 1.6. One possible outlier is Field\#2 where, for the IRIS image with the optical mask, $D=1.65$ is slightly larger than the mentioned upper limit 1.6 and is also the largest among all the fields under study. Taking into account that this field has the lowest number of successful realisations in MC and, as can be seen from Table~\ref{tab:fd_result}, the original image has a significantly lower $D\approx1.5$, we can address this outlier to an effect of optical mask influence. We discuss the impact of masking on the fractal dimension in greater detail in Sect.~\ref{sec:effects_that_change_D}.
\par
For both the Taurus and Chameleon test fields, we obtain the $D$ values which are greater than the previously measured ones: $1.36 \pm 0.06$ for Chameleon and $1.33 \pm 0.03$ for Taurus versus $1.280 \pm 0.016$ and $1.230 \pm 0.004$ in \cite{Dickman_etal1990}, respectively. However, all these measures are relatively close to each other. It is also important to note that the error budget of~\cite{Dickman_etal1990} takes into account only the error of regression fitting and, hence, is unreliably small. It is reasonable to expect that accounting for the differences, which are introduced by slightly different levels of selection or due to the filtering of small clouds (see discussion in Sect.~\ref{sec:tailcut}), should result in a larger margin of error. Thus, the results, measured here for the test images, are consistent with those from the literature. 
\par
As to the intercepts, we found that the average values are~$\langle K \rangle=1.33$ and~$\langle K \rangle=1.63$ in the optical and IR, respectively, while the spread of the values across the different fields is also considerably small: $\sigma(K)=0.09$ and $\sigma(K)=0.24$ for the optical and IR data, respectively. Our values of $K$ are smaller than those found by~\cite{Vogelaar_Wakker1994} ($1.7<K<3$, see their table~3). Perhaps, the difference in the intercepts is associated with the fact that the measured fractal dimension is, on average, slightly larger in our work, but we do not pursue this question further.
\par
Concerning the striking difference between the results in the optical and IR, there are several possible reasons for that. First of all, there is a possibility that the measured fractal dimension values somehow reflect the way the fields were processed. The major factor here is obviously the mask as it substantially changes the geometry of small clouds and also affects larger ones, although to a smaller degree. As shown by~\cite{Sanchez_etal2005}, the image resolution (i.e. how many pixels are in a cloud of a fixed size) can also change the fractal dimension value depending on the actual 3D fractal dimension. Finally, there may be some physical reasons associated with the dust properties and dynamics. We thoroughly discuss this matter in the next Sect.~\ref{sec:effects_that_change_D} and in Sect.~\ref{sec:discussion}.
\par 
Here we also note that the spread $\sigma$ of $D$ values for the clouds from different fields is rather small for both the optical and IR data. It is an interesting result since the fields are not connected in any way, see Fig.~\ref{fig:skypos}. Thus, in terms of the fractal properties, all the clouds observed seem to be similar to each other. This fact supports the idea that these cirrus clouds are close to each other in terms of the physical processes that shape them.
\par

\begin{table*}
\caption{Measured $D$ values for 8 examined optical fields and their IRIS counterparts. In addition, we include the results for \textit{Herschel} Field\#5 and two IRAS testing fields, Chameleon and Taurus. Second column lists an area of the image, columns (3) and (5) list the fractal dimensions and intercepts for the optical (or \textit{Herschel}) data; (5) and (6) lists the results for the optical data rebinned to 90 arcsec and reconvolved with the IRIS PSF; columns (7) and (8) list the results for the IRIS counterpart with the optical or \textit{Herschel} mask applied where possible, columns (9) and (10) --- for the same data without a mask. Column (11) is the difference between (5) and (7). All presented values of $D$ indicate a median for the MC realisations, the upper and lower error boundaries are 1~$\sigma$ from the corresponding percentiles.}
\label{table:main_parameters}
\begin{tabular}{c|c|c|c|c|c|c|c|c|c|c}
\hline
Field & $A$, deg$^2$ & $D$ & $K$  & $D_{90}$ & $K_{90}$ & $D_\mathrm{IRIS}$ & $K_\mathrm{IRIS}$ & $D_\mathrm{IRIS}^\mathrm{orig}$ & $K_\mathrm{IRIS}^\mathrm{orig}$ & $D_{90}-D_\mathrm{IRIS}$\\
(1) & (2) & (3) & (4) & (5) & (6) & (7) & (8) & (9) & (10) & (11)\\
\hline
\#1 & 2.0 & $1.88^{+0.02}_{-0.01}$ & 1.03 & $1.41^{+0.12}_{-0.12}$ & 1.83 & $1.49^{+0.16}_{-0.08}$ & 1.61 & $1.34^{+0.04}_{-0.05}$ & 2.02 & -0.09 \\[0.1cm]
\#2 & 0.8 & $1.66^{+0.06}_{-0.06}$ & 1.41 & $1.54^{+0.08}_{-0.04}$ & 1.49 & $1.65^{+0.09}_{-0.06}$ & 1.32 & $1.50^{+0.12}_{-0.10}$ & 1.61 & -0.10 \\[0.1cm]
\#3 & 1.8 & $1.70^{+0.04}_{-0.05}$ & 1.33 & $1.54^{+0.07}_{-0.09}$ & 1.53 & $1.59^{+0.18}_{-0.09}$ & 1.43 & $1.52^{+0.07}_{-0.06}$ & 1.53 & -0.06 \\[0.1cm]
\#4 & 2.0 & $1.66^{+0.06}_{-0.06}$ & 1.41 & $1.52^{+0.09}_{-0.09}$ & 1.53 & $1.57^{+0.08}_{-0.08}$ & 1.44 & $1.40^{+0.10}_{-0.07}$ & 1.76 & -0.05 \\[0.1cm]
\#5 & 3.0 & $1.66^{+0.05}_{-0.06}$ & 1.40 & $1.57^{+0.05}_{-0.04}$ & 1.43 & $1.43^{+0.05}_{-0.05}$ & 1.76 & $1.41^{+0.05}_{-0.05}$ & 1.75 & 0.14 \\[0.1cm]
\#16 & 1.0 & $1.64^{+0.04}_{-0.05}$ & 1.42 & $1.55^{+0.14}_{-0.07}$ & 1.46 & $1.40^{+0.04}_{-0.02}$ & 1.69 & $1.33^{+0.02}_{-0.02}$ & 2.01 & 0.15 \\[0.1cm]
\#join1 & 9.2 & $1.69^{+0.06}_{-0.05}$ & 1.31 & $1.63^{+0.05}_{-0.04}$ & 1.34 & $1.45^{+0.10}_{-0.08}$ & 1.68 & $1.35^{+0.06}_{-0.05}$ & 1.92 & 0.18 \\[0.1cm]
\#join2 & 6.8 & $1.67^{+0.05}_{-0.04}$ & 1.35 & $1.61^{+0.04}_{-0.03}$ & 1.38 & $1.32^{+0.09}_{-0.09}$ & 2.11 & $1.22^{+0.10}_{-0.05}$ & 2.35 & 0.29 \\[0.1cm]
\hline 
Avg & - & $1.69^{+0.05}_{-0.05}$ & 1.33 & $1.54^{+0.08}_{-0.09}$ & 1.50 & $1.48^{+0.10}_{-0.07}$ & 1.63 & $1.38^{+0.07}_{-0.06}$ & 1.86 & -\\[0.1cm]
\hline
\textit{Herschel} \#5 & 3.0 & $1.65^{+0.07}_{-0.04}$ & 1.37 & $1.40^{+0.08}_{-0.05}$ & 1.87 & $1.43^{+0.05}_{-0.05}$ & 1.76 & $1.41^{+0.05}_{-0.05}$ & 1.75 & -0.03 \\[0.1cm]
Chameleon & - & - & - & - & - & - & - & $1.36^{+0.02}_{-0.02}$ & 1.91 & - \\[0.1cm]
Taurus & - & - & - & - & - & - & - & $1.33^{+0.03}_{-0.03}$ & 1.98 & -\\[0.1cm]
\end{tabular}
\\
\label{tab:fd_result}
\end{table*}

\begin{table*}
    
    \caption{The results of fractal dimension measurements for cirrus clouds from our and previous studies.}
    \begin{tabular}{c|c|c|c|c|c}
    \hline
         Study & Data & $\langle D\rangle$ & $D$ range & rough field size & contours size threshold \\
         \hline
         \cite{Bazell_Desert1988} & cirrus (IRAS) & 1.26 $\pm$ 0.04 & 1.12 - 1.40 &  $\Delta \delta \sim 16^\circ$, $\Delta \alpha \sim 24^\circ$ &  187200, 8.5  \\ 
         \cite{Dickman_etal1990} &  molecular clouds (IRAS) &- & 1.174   - 1.278  & $\Delta \delta \sim 10^\circ$, $\Delta \alpha \sim 10^\circ$ &  187200, 8.5 \\
         \cite{Falgarone_etal1991} & molecular clouds (CO lines) & $1.36 \pm 0.02$ & - & $\Delta \delta \sim 10^\circ$, $\Delta \alpha \sim 10^\circ$ & -  \\
         \cite{Vogelaar_Wakker1994} &  cirrus (IRAS) & $1.42 \pm 0.02$ & 1.23-1.54  & $\sim 1  \square^\circ$ & 288000, 9.0 \\
         \cite{Sanchez_etal2005} &  Orion A (CO line) &1.35 & -  & $\Delta \delta \sim 2^\circ$, $\Delta \alpha \sim 2^\circ$ &  $\sim$777, 3.1 \\
         \cite{Juvela_etal2018} & molecular clouds (\textit{Herschel}) & $1.25 \pm 0.07$ & 1.05 - 1.4  & $\sim 0.5 \square^\circ$ &  3960, 4.7\\
         This study           & SDSS Stripe82  & $1.69 \pm 0.07$ & 1.66-1.88 & $\sim (1-9)$ $\square^\circ$ & - \\
         This study           & IRIS           & $1.38 \pm 0.09$ & 1.22-1.52 & $\sim (1-9)$ $\square^\circ$ & -\\  
         \hline
         \end{tabular}
         \\
         {\raggedright \footnotesize{\textit{Notes}: The second column lists the type of the studied objects along with the data source. Note that for~\cite{Vogelaar_Wakker1994}, we include the results only for cirrus clouds, while in the referred work, the high-velocity molecular clouds were also studied. The third column lists the average values of the fractal dimension if they are provided by the authors or, otherwise, there is a dash symbol, while the fourth column lists the range of fractal dimensions of individual clouds studied in the refereed works. The dash symbol in the fourth column means that only one cloud was studied. The fifth column lists the rough values of the overall field sizes that were studied in the cited works. The sixth column lists the values of the contours size threshold that was used to filter out small clouds in physical units (arcsec$^2$) and in units of $\log A$ that are used in the present study, respectively.}\par}
    \label{tab:previous}
\end{table*}


\par
\section{Analysis of the factors that contribute to fractal dimension measurements}
\label{sec:effects_that_change_D}

One of the results obtained in the previous section is that the measured fractal dimension of the clouds, identified in the optical data, is substantially greater than the fractal dimension of their counterparts in the IR data. At the same time, the results of fractal dimension measurements are hard to interpret from a physical point of view if we do not know how the various factors, which are incorporated in the measurement procedure (the choice of the minimal brightness contour level to account for, the choice of the lower boundary of the region size, the masking), affect the results. Our MC simulation is useful to understand more clearly how the aforementioned factors, along with some others, can affect our fractal dimension measurements.
 
\par
\subsection{Analysis of the mask influence}
\label{sec:masking}
The main obvious difference between the clouds analysed here and the ones studied in the previous studies, listed in Table~\ref{tab:previous}, is the existence of masked areas which can overlap with the structure of an actual cloud.
Therefore, one can expect that masking should somewhat change the geometry of such a cloud. The question is \textit{to what degree} it can actually change the fractal dimension of the clouds in the fields under study.
 \par
In our method, the existence of masked areas can affect a $D$ measurement in two different ways. First, the arbitrary decision, described in step (vii), to filter out all clouds, which touch the extended mask parts and are selected in step (i), can shift the resultant values of $D$ significantly. 
\cite{Bazell_Desert1988} and \cite{Dickman_etal1990} also removed from their analysis those contours which touch (or intersect) the borders of the image. However, the extended mask parts of the fields under consideration can also lie completely inside. This is another difference (see an example of such a mask in panel (b) of Fig.~\ref{fig:pipe}). 
Nevertheless, our MC analysis shows that the fractal dimension $D$ remains almost the same regardless of the decision on filtering and extended mask size selection, as illustrated in Fig.~\ref{fig:mcparamdep} for Field\#3 (parameters `mask size' and `no touch'). This is a surprising finding, since, in principle, rectangular and round blobs, which constitute the mask, can be found by chance completely inside the selected contour level, hence drastically decreasing the area $A$ while  the perimeter $P$ becomes increasing. 
\par
\begin{figure}
\includegraphics[width=0.95\columnwidth]{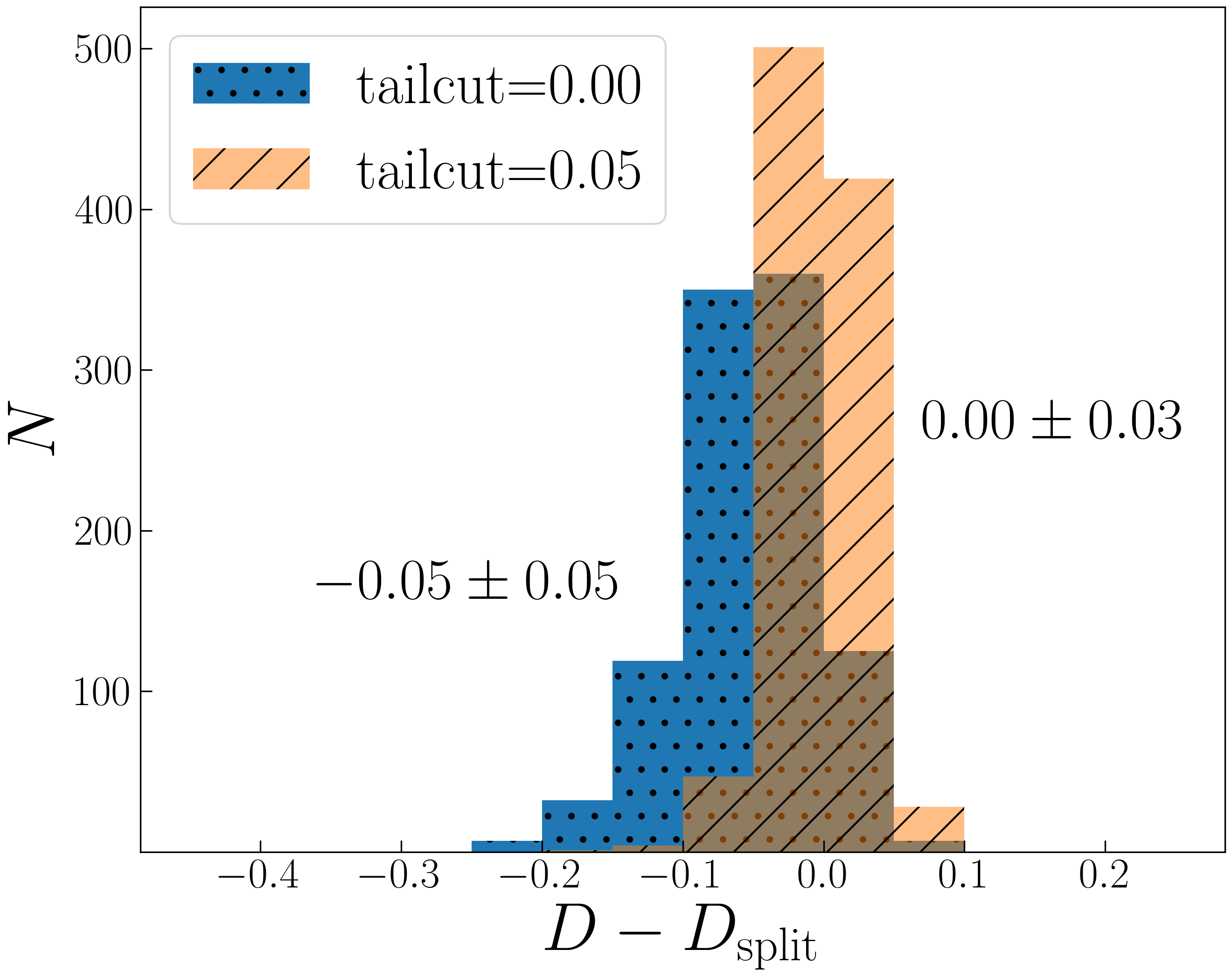}
\caption{Histograms by the $D$ variations for our experiment when individual clouds are dissected by lines (see Sect.~\ref{sec:masking} for details). The right-shifted histogram with diagonal hatching corresponds to the cases when clouds with small sizes have been removed with the `tailcut'=0.05 applied. The histogram with dots hatching represents all cases.}
\label{fig:splitsim}
\end{figure}

To study the effect when cloud regions touch the image borders or partially covered by a mask, we perform a simple experiment. First, we select more than 1500 `good' optical clouds that are not touched by extended masked regions or image borders from all fields and different contour levels. Then we randomly select some small number of clouds\footnote{This choice is motivated by the inner plot in Fig.~\ref{fig:lev_dep}, where the median of the natural logarithm $\log N \approx 4.5$, and, thus, $N \approx 90-100$ can be used as some characteristic number of clouds in a realisation, on average.} $n<100$ and measure the fractal dimension $D$ for them. After that, we dissect $m < n$ of these clouds by a straight line, whose slope and intercept was chosen randomly. The resulting pieces of the original clouds are then treated as new clouds. A set of such clouds, along with the remaining unsplit clouds, are characterised by a new fractal dimension $D_\mathrm{split}$. This process reliably simulates the situation when cirrus clouds are ``spoiled'' by the image boundaries or a mask. The values of $D-D_\mathrm{split}$ for 1000 realisations are shown in Fig.~\ref{fig:splitsim}. This figure can be used to determine how the effect of masking impacts the measurements. On average, the splitting into smaller clouds yields $D_\mathrm{split}$ which is larger by 0.05 than the original one. This can be explained by the fact that the number of small noisy clouds grows after the slicing. If we filter out those small clouds using the threshold `tailcut'=0.05, the resultant histogram by $D-D_\mathrm{split}$ becomes symmetric and centred at around zero with a standard deviation of just 0.03. These errors are too small to change the overall result of $D$ measurements. This answers the main question of this subsection.
\par 
The second effect of masking in the optical relates to the ``holes'' it can produce in the interior of the clouds. As was already mentioned, if a data satisfies a condition $P \propto A^{D/2}$ and we apply an internal mask to it, then the resulting set of points $(\log A, \log P)$ will have a larger $D$. In our MC analysis, we can either interpolate such a mask in step (ii) before selecting a contour or fill the ``holes'' after the contour selection in step (v). Both these steps have almost no impact on $D$ estimation as can be seen from Fig.~\ref{fig:mcparamdep} (the parameters `interp' and `fill holes'). The reason behind this is that the ``holes'' are usually less than 5 pix wide and they can significantly change the position for only small clouds in the $\log A \div \log P$ plane, which are filtered out in most of the cases anyway. However, this does not hold true for the IRIS counterparts and rebinned optical fields, because due to lower resolution the sizes in pixels of all clouds in them are smaller than those in the original optical images. To verify this, we apply the mask, which we use for the optical data, to the IR images and carry out a MC analysis. We set a pixel masked only if more than half of the original (small) pixels within the rebinning window are masked. As an illustration, one can check the resulting mask for Field\#5 in panel (d) of Fig.~\ref{fig:f5_comp} and compare it with the original image in panel (a). It is clear from Fig.~\ref{fig:allmc3}, panel (b), that, on average, the fractal dimension $D$ is larger for the IRIS images with the mask applied (grey) than without it (magenta). We elaborate more on this result in panel (d) of the same figure, where we additionally show the results for the MC realisations with the filled ``holes'' only. As one can see, filling the ``holes'' lowers the $D$ value to the level which was measured before the masking. We note that these $D$ values are not entirely identical because the mask can also filter out some cirrus clouds, as discussed in the previous paragraph. Nevertheless, even if we can clearly see that the mask increases $D$ values for a low-resolution data, these values are shifted by less than $\approx0.1$ and still remain within the same margin of error. It is worth noting here that masking was rarely done in the previous studies. However, masking is also important for IR wavelengths because the cirrus emission can be contaminated by non-dust sources such as UltraLuminous InfraRed Galaxies (ULIRGs). This makes the obtained results valuable for future studies at different wavelengths.

\subsection{`Tailcut' parameter}
\label{sec:tailcut}
\begin{figure}
\includegraphics[width=0.95\columnwidth]{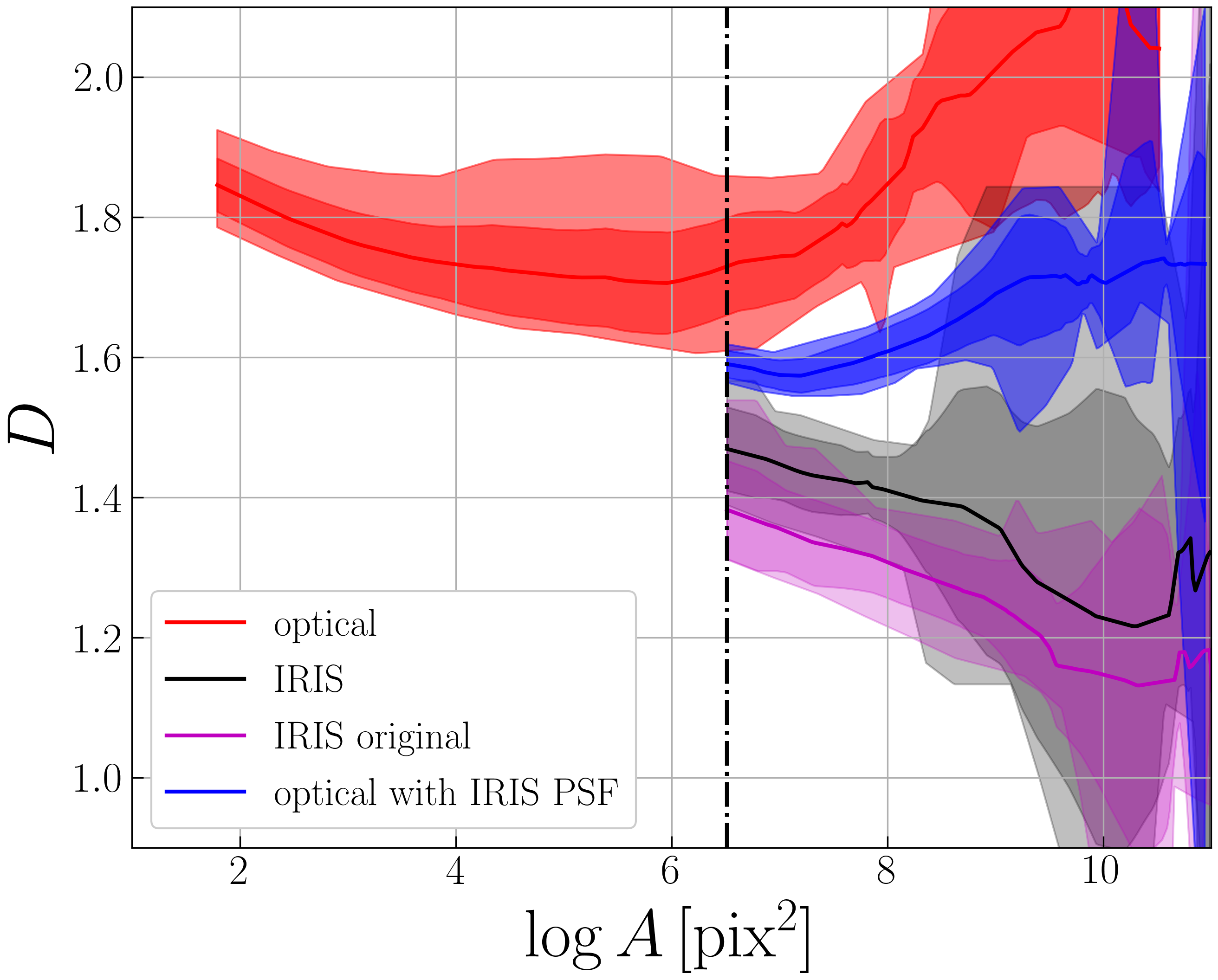}
\caption{Dependence of $D$ across all fields on threshold to filter out small clouds (i.e. `tailcut' parameter in each realisation). The colours represent individual sets of data, the solid line for each set depicts an average curve, the inner colour-filled spread corresponds to one standard deviation, whereas the outer spread shows the minimal and maximal values. The vertical depicts the 3~square pix size for the IRIS data resolution as a minimal used threshold for the filtered area.}
\label{fig:mc}
\end{figure}

As mentioned in Sect.~\ref{sec:methods}, the lower boundary for the size of regions, we account for in linear regression fits, is parametrised by the `tailcut' parameter.  
In~\cite{Bazell_Desert1988} and~\cite{Dickman_etal1990}, a typical size of the regions excluded from consideration was less than 52 arcmin$^2$~(13 square pix of the original IRIS resolution). In~\cite{Vogelaar_Wakker1994}, the authors found that depending on the choice of the lower limit for the region size, the results of a fractal dimension measurement will be different and that there are some ``stability regions'' in which the fractal dimension is independent of the selected value for region size. In subsequent studies, e.g.~\cite{Juvela_etal2018}, the excluded regions were much smaller in size (1.1 arcmin$^2$ or about 3 square pixels in their study). Our simulations allow us to find the exact way the fractal dimension depends on the selected value of the tailcut parameter. Note that this parameter is, by definition, dimensionless because it is a fraction from 0.0 to 1.0 where lower and upper bounds correspond to \textit{min} and \textit{max} of $\log A$, respectively. For any particular realisation these bounds are different and determined during the field processing. Thus, in each realisation one can indeed translate the `tailcut' fraction into some number of square pixels. Fig.~\ref{fig:mc} shows how the measured fractal dimension of the clouds changes with the value of the lower boundary size of the selected regions for the optical and IR data. For ease of presentation, we averaged the dependencies over all of our fields. As can be seen from the figure, for small values of the `tailcut' parameter, there is a plateau for the optical data over which the fractal dimension does not change much. After some threshold $\log A > 8$ for both the optical and IR data, the fractal dimension changes notably: it increases for the optical data and decreases in the IR. In general, we can conclude that the fractal dimension $D$ does depend on the size of the regions one decides to include in a fit. Due to the significantly better resolution of the optical data, there is a large ``stability region'' where one can safely measure the fractal dimension. The existence of such dependence is particularly important for IR data where one cannot find a similar stability region. 
\par
\begin{figure}
\includegraphics[width=0.95\columnwidth]{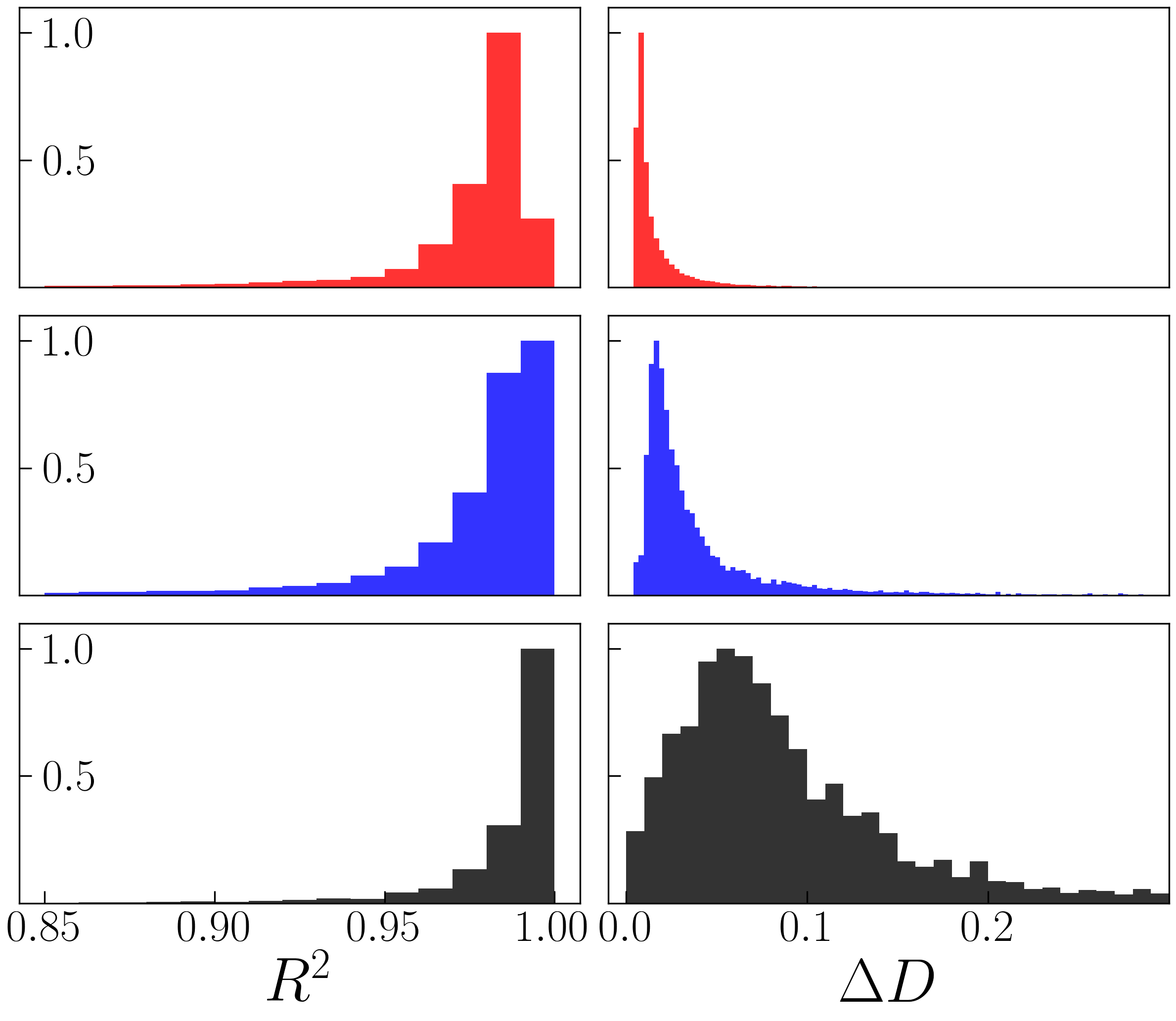}
\caption{The left panels show histograms for the coefficient of determination $R^2$ of the fitted MC regressions. The right panels show distributions of the regression slope fitting error obtained using an ordinary least squares method and multiplied by two, thus equal to the $D$ estimation uncertainty. The upper panels show the optical data, the second row is for the optical data convolved with the IRIS {PSF} and the bottom one is for the IRIS data. All histograms are normalized to unity size of the maximal bar for better visibility.}
\label{fig:linregerrors}
\end{figure}

Another small but important addition here is that a large value of the `tailcut' parameter can significantly affect the quality of $\log P \div \log A$ regression fitting. It is easy to imagine a situation when there are only a few points left and $D$ is not well constrained from such a limited data. Of course, this is not the only parameter responsible for such cases but the most important one. The quality of the MC regression lines is illustrated in Fig.~\ref{fig:linregerrors}. It is easy to see that, as the coefficient of determination $R^2$ suggests the total variation explained by the linear regression, it appeared to be lower for the optical data. The number of points is larger there, and, for exactly the same reason, the line fitting errors are greater for the IR fields. Nevertheless, these errors are small and the overall quality of the regression fits is good. 

\subsection{Image resolution and PSF}
\label{sec:psf}
The image resolution also affects the fractal dimension since the area and the perimeter are measured using discrete pixels. Therefore, the more pixels are contained in some selected cloud, the more accurate the estimate of its perimeter and area will be. Clearly, as the IR data has 15 times lower image resolution than the SDSS optical data we use, the image resolution can be a reason why the resultant fractal dimension is quite different for these two. To measure the effect of image resolution on fractal dimension measurements, we performed a simple test. Specifically, we rebin the optical data to the image resolution (pixel size) of the IR data. The intensity of the large resulting pixels is obtained by summing up the intensities of the original small pixels. The resultant values of the fractal dimension obtained from such reduced images are shown in Fig.~\ref{fig:allmc3}, panel (c) (light blue points). {As can be seen, the effect of the resolution appears to be  almost negligible: the fractal dimension values only shift by $~0.05$, on average, to the lower end across the different fields.} This is comparable with the $D$ measurement error. In~\cite{Sanchez_etal2005}, an equally small shift in the values after decreasing the image resolution was observed in case of rather low volumetric fractal dimension $D_\mathrm{vol} \sim 1.2-2.0$ (see their figure 7). Perhaps our results indicate that we indeed deal with such clouds, but we cannot rule out the possibility that for real clouds the dependence of the projected $D$ on the volumetric $D_\mathrm{vol}$ and image resolution should be more complicated than that obtained by~\cite{Sanchez_etal2005} on the example of some simulated clouds.
\par
Apart from the image resolution, there is another important factor that distinguishes our IR and optical data, namely, the angular resolution, or the PSF. The effect of the PSF on fractal dimension measurements is hard to predict, but a larger PSF blurs out intensity gradients more effectively, and, therefore, leads to smoother contours of objects, in general. To estimate the consequences of the PSF differences for our optical and IR data, we applied the IRAS PSF with FWHM=4.3~arcmin to the optical data with the artificially lowered image resolution described in the previous paragraph (in other words, we convolved the optical data to the IR one). There is no need to consider the optical PSF here since the entire optical PSF is contained within less than one pixel of an IRIS image. The resultant fractal dimensions for all fields are shown in Fig.~\ref{fig:allmc3}, panel (c), and listed in Table~\ref{tab:fd_result}. It can be seen that the fractal dimension is actually quite dependent on the PSF size and it becomes significantly smaller (on average) for the degraded optical data as compared to the initial one. For Fields\#1-5, the fractal dimension of the clouds, observed in such degraded optical images, is now consistent with the results for the corresponding clouds in the IR~(see Table~\ref{tab:fd_result}, $D_{90}-D_{IRIS}$ column). After accounting for the differences in the PSF, a significant disagreement between the fractal dimensions for the degraded optical data and IR fields is essentially maintained for our largest fields, Field\#join1 and \#join2, and for Field\#16. For these fields, most likely, we trace different structures in the IR and optical due to the large-scale background subtraction in the SDSS (see Discussion). However, note that for these fields the differences between the fractal dimensions in the optical and IR become smaller too after convolution with IRIS PSF. To our knowledge, the influence of the PSF has not been previously considered in the context of fractal dimension measurements, but, apparently, it is one of the most important determining factors for measuring the fractal dimension of cirrus clouds. Therefore, this finding is of great importance for future studies aimed at exploration of the geometric properties of dust clouds in our Galaxy.
%
\par
It is important to note that both findings presented above are also valid for the \textit{Herschel} Field\#5 data, as can be seen for the green points in Fig.~\ref{fig:allmc3}. There is also no need to consider the \textit{Herschel} PSF for the above reason. Taking into account that the \textit{Herschel} $250\mu$m image corresponds to a slightly different part of the light spectrum as compared to {\it IRIS} 100~$\mu$m, but nonetheless, its analysis yields the same result. This makes our conclusion even more robust and emphasises the importance of the correct PSF conversion for a proper comparison between different data.

\subsection{Brightness contour level}
In~\cite{Bazell_Desert1988}, the authors calculated the fractal dimension of the clouds for different levels of brightness separately, while \cite{Dickman_etal1990} fitted a linear regression to the points obtained for a set of selected levels. They verified that their results remained the same if they considered only one specific level, although in that case the errors appeared larger. \citet{Vogelaar_Wakker1994} studied how the fractal dimension changes with the lowest contour level selected and found that the dimension of some clouds can depend on it.
\par 
In the present work, we use our MC simulation to explore how the choice of the brightness contour level to distinguish cirrus clouds affects the resultant fractal dimension value. Fig.~\ref{fig:lev_dep} showcases how the fractal dimension value depends on the brightness level for all of the optical fields processed in the same way, i.e. they represent some subsample of MC realisations where all parameters remained the same. As can be seen, the choice of the brightness level indeed affects the resultant value, although to a small degree. On average, there is some linear trend with a higher brightness level corresponding to smaller values of the fractal dimension. The latter statement is also illustrated in Fig.~\ref{fig:mcparamdep}, the first panel in the second row. The reason why brighter contours resulted in a lower $D$ can be seen in Fig.~\ref{fig:pipe}, panel (e). With an increase of brightness, clouds become smaller, on average, and the left tail of the size distribution (shown by red points) starts to influence the result harder, while for such points the perimeter $P$ decreases faster than the area $A$. To be more confident in our results, we checked how the results, presented in Fig.~\ref{fig:allmc3}, would change if we included the structures, which are distinguished using all brightness level contours in one regression fit (see Sect.~\ref{sec:discussion} for more details). We plot the values obtained in this way in Fig.~\ref{fig:allmc3} --- depicted by horizontal segments of the same colour as the underlying data. It can be seen that accounting for all brightness levels in one fit does not change the values of the fractal dimension nor does it decrease the difference between the IR and optical data.

\subsection{Colour filtering}
\label{sec:color}
The results of~\cite{Roman_etal2020} suggest that cirrus clouds can be distinguished using optical colours~(see Sect.~\ref{sec:methods}, step (iv)). We apply this criterion for a part of the realisations in our MC simulation and check if filtering the pixels by their $(r-i)$ and $(g-r)$ optical colours has an impact on estimation of $D$.  As can be seen from cyan points in Fig.~\ref{fig:allmc3}, the actual difference between the two sets of values is significant and approximately equals 0.2 for different fields. The colour filtering step shifts $D$ toward larger values, because it can produce a more perforated structure with additional holes, which leads to an overall increase of the $D$ value (see clarifications in Sect.~\ref{sec:masking}). Such behaviour is also illustrated in Fig.~\ref{fig:mcparamdep}, the third panel in the first row, where the two peaks corresponding to the realisations with and without colour filtering are clearly shifted by about 0.2 with respect to each other. For better understanding of how the procedure of colour filtering change the shape of the actual clouds, we refer the interested reader to figures 13-15 in \citet{Roman_etal2020}, where authors apply aforementioned criterion to images from Hyper Suprime-Cam Subaru Strategic Program data \citep{Aihara2018}. It is clear from those images that the colour filtering transforms the clouds into much less connected structures. Moreover, the resulted ``holes'' are very different from those produced by masking in size and number.
\par
Overall, the presented results show that the colour filtering has a strong impact on the fractal dimension measurements. 
However, we should emphasise that there is room for some doubt as to whether this result really reflects the physical properties of the clouds under consideration. And, particularly, for the following reasons.
First, the calculation of $D$ through colour filtering has not been done before, thus we cannot compare our findings with previous works. Secondly, we do not know how good is the filtering relation in statistical terms. In fact, a given pixel could be outside the selected colour constraints due to low signal-to-noise ratio, but this does not mean that this pixel does not contain dust emission (see left subplot in fig.~12 in \citealt{Roman_etal2020}, where some grey error bars lie above the relation line). Probably, this affects clouds at the lowest surface brightness only, but this needs to be tested additionally. Thirdly, as mentioned earlier, the shape of the selected clouds after filtering differs significantly from those to which we do not apply the colour criterion (compare the clouds presented in Fig.~\ref{fig:lev_dep} and Fig.~\ref{fig:pipe} and those in figures 13-15 in \citealt{Roman_etal2020}). It is clear that the methods we use in our MC simulation will not help to solve connectivity and ``holes'' issues in such differently shaped clouds. Probably, an additional rebinning step can improve the situation by smoothing the statistical fluctuations, but this can lead to new biases and needs to be tested anyway. Taking into account all the reasons described above, we decided to not include the realisations with colour filtering into our final $D$ estimation and its error margins. Nevertheless, the result obtained here has a methodological importance, which, along with the colour relation itself, definitely deserves attention and an additional investigation in a separate study.

\par
The only case where the filtering has almost no effect at all is Field\#1, where the $D$ value is anomalously large and inconsistent with the others. Also, it is very stable, that is it shows almost no estimation error. This is even stranger since Field\#1 has the lowest $<g-r>$ colour among all, as shown in table~2 from \cite{Roman_etal2020}. Therefore, it has a more strict upper condition which should filter out, on average, more pixels and, thus, the filtering should affect the result even more. The reason for such a strange behaviour is that Field\#1 contains clouds of a significantly smaller size. This means that the point with the largest area $A$ of Field\#1 has an area that is more than an order of magnitude smaller as compared to the other fields. This addresses the issue under consideration since the colour filtering should affect large clouds the most.
\par
It is also interesting to note from Fig.~\ref{fig:allmc3} that $D$ values after the colour filtering are well consistent with each other between different fields, but whether this is an artificial result or not remains an open question. As already mentioned above, this is a potentially large and completely new area that requires further research.

\subsection{Different bands}
\label{sec:diffbands}
The values of the fractal dimension obtained for the optical data should be averaged over four available optical bands: $g,r,i,$ and $z$. However, for the individual bands, the values can be slightly different, so we should verify how large this difference is. The value of the fractal dimension $D$, averaged over all fields and realisations, is equal to $1.73, 1.74, 1.77$, and $1.89$, accordingly, in the same order as above and with a typical error of around 0.08. As can be seen, for all bands save the $z$ band, the values are somewhat consistent with each other, while for the $z$ band the average value is larger by about 0.15. The true reason for such a behaviour is hard to pinpoint accurately, but we should note that SDSS imaging in the $z$ band is substantially shallower than in the other bands, that is, the $z$ band surface brightness limit is $\mu_\mathrm{limit}\sim26.6$~mag arcsec$^{-2}$, as compared to $\mu_\mathrm{limit}\sim28.5$~mag arcsec$^{-2}$ \citep{Roman_etal2020}, on average, for the other bands. Consequently, there should be a smaller number of dim clouds in the $z$ band, which can be distinguished in principle. Perhaps, this circumstance leads to the observed increase of the fractal dimension in the $z$ band. We can also see by visual inspection that the data in this band have more artefacts and issues, especially where the original SDSS frames were composed. Since results in the $z$ band are inconsistent with those obtained for the other SDSS bands, we decided to exclude this band from our final results averaged over all bands under study (i.e. reported in Table~\ref{tab:fd_result} and shown in all figures).
\par
{\,}
\par
Let us briefly sum up the results of this section. Most of the results are illustrated in Fig.~\ref{fig:allmc3}. We have found that masking affects the results in two different ways. It can artificially increase $D$ values by ``making holes'' in the clouds. Secondly, extended mask parts and image boundaries can obscure or even split up individual clouds. Both these effects are found to be small ($\Delta D < 0.1$). Unlike masking, filtering of small clouds with the `tailcut' parameter have a great impact on $D$ estimation. We show that the optical data demonstrates the same fractal dimension within some `plateau', while for the IR counterparts, $D$ can only decrease. A change of image resolution only slightly reduces $D$, but convolution of the optical data with the IRAS PSF lowers the fractal dimension significantly. The obtained results are almost the same for different brightness contour levels, as expected for fractals. The calculation of $D$ through colour filtering was done for the first time in this work, and we found that it can significantly increase $D$ by a value around 0.2. This effect can be artificial due to several reasons listed above, and it needs further investigation. Finally, the $z$ band was found to be inconsistent with the other SDSS bands and was excluded from our final results. {Note that the cumulative effect of all aforementioned factors on $D$ should not be considered as a sum of individual factors because the correct uncertainties on $D$ have been specially estimated in our MC simulation.}

\section{Discussion}
\label{sec:discussion}

\begin{figure}
\includegraphics[width=0.95\columnwidth]{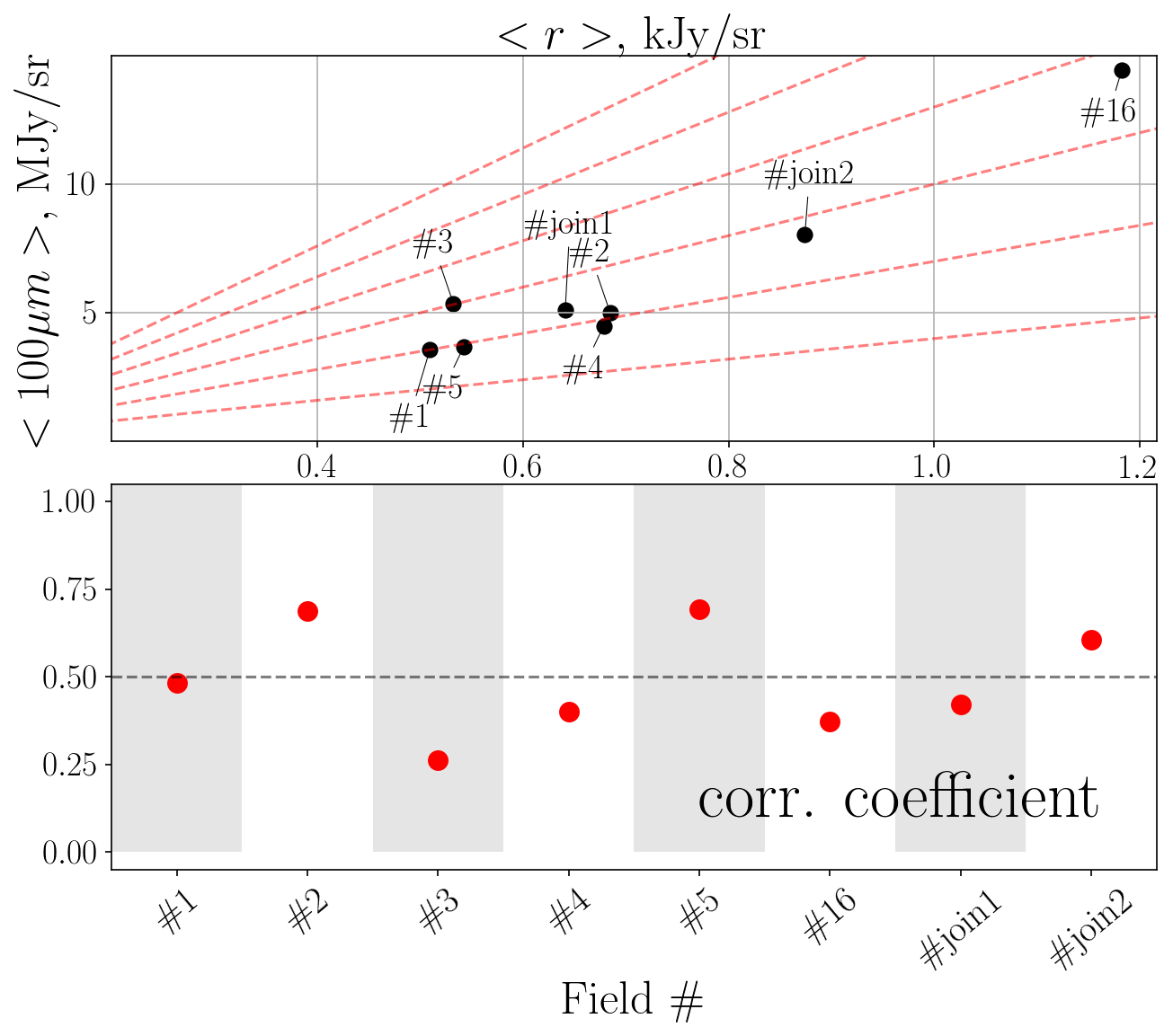}
\caption{Top panel shows average pixel fluxes in the $r$-band fields and their IRIS 100~$\mu$m counterparts. All images are rebinned to the same image and angular resolution, and processed with an identical optical mask. The red dashed lines, which start in the origin of the coordinates, show linear dependencies with different slopes. The bottom panel shows the pixel-by-pixel Pearson correlation coefficient for the same data.}
\label{fig:meanflux}
\end{figure}



%
\subsection{Optical versus IR fractal dimension}
The analysis, which we carried out in the previous section, convincingly shows that if we account for various differences between optical and IR data and the differences in the corresponding data processing, there are some fields (Fields\#5-6, \#join1, \#join2) where the fractal dimension values in the optical are considerably greater than those in the IR. For the other fields (Fields\#1-4), the measured fractal dimensions are quite close to each other. Below we discuss the possible reasons leading to the observed consistencies and inconsistencies.
\par 
First of all, we assume that the light comes from the same dust grains in the optical and IR. However, different parts of their spectrum we observe are mainly associated with different generating mechanisms. In the case of the IR, it is a thermal emission while it is a scattering of incident light in the optical~\citep{Draine_2003}. The cirrus clouds are usually optically thin~\citep{Stark1995,Szomoru_Guhathakurta1999} and, hence, the flux should be proportional to the column density in both cases. However, there are various factors that can reduce the correlation between the flux and column density. For the thermal emission, there are variations in the dust temperature~\citep{Lehtinen_etal2007,Ienaka_etal2013}. For optical fluxes, it is important from where the photon comes and how it is scattered, i.e. we should know the phase function. 
\cite{Seon_Witt2013} verified how the scattered flux correlates with the optical depth for different values of the phase factor $g$ and homogeneous optical depth values on the example of some simulated clouds where the source of radiation is located in the centre of the cloud. They found that there is a good correlation between the optical depth and the scattered fluxes maps in the case of complete forward scattering, i.e. $g=0.99$~(see their figure 3). However~\cite{Seon_Witt2013} also found that decreasing the $g$ factor (i.e. the case of more isotropic scattering) decreases the correlation between the fluxes and optical depth. For cirrus clouds, which are illuminated by the interstellar radiation field (and not a point source as considered by~\citealt{Seon_Witt2013}), we can assume that the scattering picture we observe should be closer to the case of isotropic scattering described by~\cite{Seon_Witt2013}. Nevertheless, from their figure~1 we can see that the correlation is still strong even in the case of $g\lesssim 0.5$. For the typical value of $g \sim 0.6-0.7$~\citep{Ienaka_etal2013}, the results of~\cite{Seon_Witt2013} suggest that if we measure some geometric property (fractal dimension) of the clouds in the optical, it can have a different value than if measured based on IR data (which is a better tracer of the dust density). 
   
\par 
These and several other reasons can lead to a situation that we indeed measure the fractal characteristics of objects, the images of which are different in some particular details at IR and optical wavelengths. This is especially true for some of the fields under consideration. This fact can be noticeable, for example, from the one-to-one visual comparison of lower panels in Fig.~\ref{fig:f5_comp} and is illustrated in Fig.~\ref{fig:meanflux}. The mean pixel fluxes in the optical and IR are compared in the upper panel. It can be seen that an IR emission increase does not always coincide with a significant optical intensity change. The same conclusion can be drawn based on the lower panel, where the pixel-by-pixel correlation coefficient is measured for the IR and optical fluxes. In only three cases, we found that the data behave in the same way, while in all other cases the correlation is weak. It is clear for our data that in the fields under consideration a direct relation between the images of the cirrus clouds, if exists, is at least nonlinear. In general, this is not a new result and has been known for a long time \citep{Guhathakurta1989}.
\par
Besides the underlying physical factors mentioned above, in our case, there are several additional reasons for the observed discrepancy between the IR and optical structures. First, it is partially due to the presence of bright unmasked sources such as, for example, ULIRG F22509-0040 in Field\#4. However, such sources are few. The second reason is large-scale subtraction of the background in the optical fields as an inevitable drawback of the SDSS data reduction pipeline. The background subtraction procedure implies a polynomial or spline approximation of the background and removes every diffuse emission with a scale larger than several arcmins (see section 4 in \citealt{Roman_etal2020}). As the result of this process, extended details of cirrus clouds (or large-scale diffuse emission) can potentially be removed from our optical images and, thus, not included in the $D$ evaluation. However, as we have already demonstrated with high confidence, the constancy of the fractal dimension value between different contours, as well as different parts of Stripe82, is great and, thus, it is reasonable to presume that, at least on small scales, it should be reliably measured and nonaffected by possible over-subtraction. This point can be additionally illustrated by the fact that one of the largest fields Field\#join2 demonstrates a strong correlation between optical and IR cirrus parts, as Fig.~\ref{fig:meanflux} suggests. It can be concluded that the issue about the correspondence of optical and IR data is a difficult one and needs additional research, which we are about to do in a separate study.
\par 
We should note that, for dense molecular clouds with grains of sizes from $0.1$~$\mu$m to $10$~$\mu$m, there is also an interesting mechanism that can introduce the differences in the fractal dimension for the optical and IR data. It is known that the parameters of grain size distribution determine the extinction curve behaviour, as well as the scattering properties of a particular cloud~\citep{Weingartner_Draine2001} and, thus, govern the overall intensity level at different wavelengths. At the same time, based on the results of hydrodynamic simulations, various authors pointed out that for real observed clouds the larger grains can decouple from gas flows and tend to occupy denser areas, while small grains are typically well-coupled with the gas flows \citep{Hopkins_Lee2016,Tricco_etal2017,Mattsson_etal2019}. In terms of the fractal dimension, the smaller value of it means that, in general, the cloud has a more filamentary wispy structure. Hence, we found that some clouds in the IR demonstrate a more filamentary structure as compared to the optical. Qualitatively, this result agrees well with the results of~\cite{Hopkins_Lee2016}, \cite{Tricco_etal2017}, and \cite{Mattsson_etal2019}. We expect that larger dust grains, which are usually traced in the visible, inhabit denser parts of the clouds while small grains are better traced in low density filaments, which become more apparent in the IR. Although, we should admit that for a diffuse medium such as in a cirrus cloud, we expect that there is no such a wide spread by the grain size as considered by the mentioned authors. Further physical simulations of dust clouds are required to support this hypothesis.

\subsection{Comparison with previous studies}
\label{sec:comparison}
%
From Table~\ref{tab:previous}, it is interesting to note that in some earlier works~\cite{Bazell_Desert1988,Dickman_etal1990}, the dimension measured using IR data seem to be slightly smaller, on average, as compared to that obtained by~\cite{Vogelaar_Wakker1994} and the average value $\langle D\rangle=1.38^{+0.07}_{-0.06}$ in the present study for the IR data. Here we should also emphasise that our IR data have exactly the same image and angular resolution (PSF) as in~\cite{Bazell_Desert1988, Vogelaar_Wakker1994} and, therefore, some other factors should contribute to the observed inconsistency.

Based on the analysis presented in Sect.~\ref{sec:effects_that_change_D}, one can think of several possible reasons for that. First, we found that the fractal dimension value {depends on the initial filtering by the cloud size}. For the IR data, the exact value of this border seem to be one of the determining parameters. In Table~\ref{tab:previous}, we collected the values of exact sizes of the filtered regions used in the literature, where possible. As can be seen, in earlier works the authors filtered the clouds of rather large sizes. As we showed in Fig.~\ref{fig:mc}, such filtering indirectly leads to lower values of the fractal dimension $D$. The second reason is connected to the first one. As have been already noted, {smaller} clouds tend to have a smaller ratio of the perimeter and area logarithms than the larger ones. At this point, it is important to know how large the field size under study is and, according to that, how large, on average, the clouds under study are. Table~\ref{tab:previous} lists the rough values of the field sizes which were used for $D$ estimation in the previous studies. As can be seen, in earlier works, the authors used significantly larger fields (except for \citealt{Vogelaar_Wakker1994}, the results of which are most consistent with ours) than we analyse in this study. From Fig.~\ref{fig:mc}, one can expect that an increase {of the number of} of large clouds should decrease the fractal dimension.

One potentially major source of difference between this study and the previous ones is the way to choose which data to fit regression on. In our approach only one contour per realisation was used for this purpose, while in other studies starting from \cite{Dickman_etal1990} a set of shrinking contours were used with some increasing brightness interval between them. Since the interval choice is arbitrary, one needs to carry out a proper investigation of its effect on $D$. On the other side, the use of many contours at once should produce a more stable result because the same cloud is included several times and the points are placed more evenly in the $A$ range. In MC, the number of points is smaller and there is a chance that, for example, we pick just the brightest cores and do not fit the regression well due to a short data range. In order to examine if the derived results will be the same in case we use several contours instead, we carried out the following experiment. For a given Field all MC realisations were placed on the same $\log A \div \log P$ plane and then the $D$ value was measured, thus emulating a use of a huge set of different evenly distributed contours. We also use the conservative `tailcut' value 0.05 for filtering out the noisiest small clouds from the fitting. The results are presented in Fig.~\ref{fig:allmc3} by horizontal segments, where we compare the $D$ values from MC and from the described experiment. We note that the errors of the regression fitting with this number of points {are of the order of 0.01 and, thus, not shown.} For both the optical and IR data, in all cases with just a few exceptions we obtain that $D$ values lie within the uncertainty range of MC simulation and, often, they are close to its centre. Several exceptions are related to images with small resolution and can be addressed to situation when the numbers of clouds and successful realisations are small. Overall, the agreement is good and it is unlikely to be a source of disagreement in fractal dimension estimation. Note also that in the described experiment there is no tendency to produce a consistently greater or lower $D$ than MC shows.

\par

\section{Summary}
\label{sec:sum}

In the present study, we have carried out a comparative analysis of the \textcolor{black}{2D} fractal dimension $D$ for cirrus clouds in our Galaxy using IR and optical data. To the best of our knowledge, for the optical data, such a study has been conducted for the first time. We have considered 8 fields from SDSS Stripe82 which were used by~\cite{Roman_etal2020} to study the colour characteristics of the cirrus clouds in the optical. The corresponding IR counterparts have been taken from the IRIS 100~$\mu$m database. For one optical field, we have also considered the corresponding \textit{Herschel} 250$\mu$m counterpart. The deep optical fields and the mentioned \textit{Herschel} field were originally prepared by~\cite{Roman_etal2020} with accounting for the instrumental scattered light and masking of all the external sources, exposing only the diffuse emission.
\par
We have used a simple approach to compute the fractal dimension of the cirrus clouds based on the perimeter-area ratio of the cirrus contours. One of the aims was to compare our results for the IR  with the results from the literature where the same approach was used. For each of the optical and IR fields, we distinguished the structures which are enclosed by some brightness contour level, calculated the corresponding perimeter $P$ and area $A$, taking into account the {cloud boundaries and possible masked inner pixels} for each of such structures, and then approximated the $\log P \div \log A$ dependence with a straight line. The angular coefficient of the line gives us the value of the fractal dimension for the {chosen} brightness level of the cirrus clouds observed in each field.
\par
In previous studies, the error on the obtained fractal dimension was assumed to be equal to the error of a linear regression fitting. In the present paper, we performed a MC simulation to estimate how the choice of some subjective parameters (e.g., the lower size limit of clouds to be filtered out, the brightness level to choose) affects the results and, based on that, estimated an error value on $D$.  

\par
For IR data, we found that the average fractal dimension across all fields is $\langle D \rangle = 1.38^{+0.07}_{-0.06}$ with $\sigma(D)=0.09$. The obtained values are generally consistent with the results of previous studies~\citep{Bazell_Desert1988,Dickman_etal1990,Falgarone_etal1991,Vogelaar_Wakker1994,Hetem_Lepine1993,Sanchez_etal2005,Juvela_etal2018}. For our data, we found that there is a strong dependence of the fractal dimension value on the subjective parameters and, especially, on the lower size limit of the clouds to be filtered out. Accounting for this effect in our MC simulation yielded the significantly greater error $\sigma(D)$ as compared to those obtained in the literature.
\par 
In the optical, we found that the average fractal dimension of the clouds is $D=1.69$ with the very small $\sigma(D)=0.02$ if one outlier is excluded. Consequently, the $D$ values in the optical appeared to to be considerably greater than those obtained for the IR data. We explored whether this difference arises due to the differences in the observational characteristics of the field under study, namely, the image resolution and the PSF properties. We came to conclusion that the PSF has a substantial effect on the fractal dimension while the image resolution itself does not significantly affect the result. Our finding that the PSF can significantly influence the measured fractal dimension is important from a methodological point of view: one should compare the results of fractal dimension measurements for different data (and simulations) only if they have been convolved to the same angular resolution. 
\par
Concerning the effect of masking, based on our MC simulation and tests with the IRIS data, we found that the masking itself leads to an overall increase of $D$ by about 0.1, but it can be greater than that in some cases depending on data studied. As an example of such an exceptional case, the cirrus clouds in Field\#2 demonstrate an increase of the fractal dimension by about 0.15 after applying the optical mask.
\par 
As to the threshold on the contour size to filter out small clouds, which is discussed, for example, in~\cite{Vogelaar_Wakker1994}, we found that for our IR data there is a strong dependence of the fractal dimension {on this parameter}, while for the optical data there is a plateau where the fractal dimension can be reliably measured.  
\par 
In total, we conclude that if we take into account the differences in the PSF, image resolution, and masking, half of the fields under study  demonstrate equal fractal dimensions of the selected clouds in the optical and IR. For the other half, the fractal dimensions in the optical and IR remain inconsistent. We discussed various reasons for the observed phenomenon including the differences in flux generating mechanism at different wavelengths, presence of some bright unmasked sources in IR data, and the tendency of larger grains to decouple from smaller ones, proven in some model studies. We conclude that none of these reasons can be considered as dominant and additional studies are required to explain why the fractal dimensions of IR and optical cirrus can be essentially the same for some fields, while for others they can differ significantly.

\section*{Acknowledgements}
We acknowledge financial support from the Russian Science Foundation (grant no. 20-72-10052).
\par
We thank the anonymous referee for his/her
review and appreciate the comments, which contributed to improving the
quality of the article.
\par

Funding for the Sloan Digital Sky 
Survey IV has been provided by the 
Alfred P. Sloan Foundation, the U.S. 
Department of Energy Office of 
Science, and the Participating 
Institutions. 
\par
SDSS-IV acknowledges support and 
resources from the Center for High 
Performance Computing  at the 
University of Utah. The SDSS 
website is www.sdss.org.
\par
SDSS-IV is managed by the 
Astrophysical Research Consortium 
for the Participating Institutions 
of the SDSS Collaboration including 
the Brazilian Participation Group, 
the Carnegie Institution for Science, 
Carnegie Mellon University, Center for 
Astrophysics | Harvard \& 
Smithsonian, the Chilean Participation 
Group, the French Participation Group, 
Instituto de Astrof\'isica de 
Canarias, The Johns Hopkins 
University, Kavli Institute for the 
Physics and Mathematics of the 
Universe (IPMU) / University of 
Tokyo, the Korean Participation Group, 
Lawrence Berkeley National Laboratory, 
Leibniz Institut f\"ur Astrophysik 
Potsdam (AIP),  Max-Planck-Institut 
f\"ur Astronomie (MPIA Heidelberg), 
Max-Planck-Institut f\"ur 
Astrophysik (MPA Garching), 
Max-Planck-Institut f\"ur 
Extraterrestrische Physik (MPE), 
National Astronomical Observatories of 
China, New Mexico State University, 
New York University, University of 
Notre Dame, Observat\'ario 
Nacional / MCTI, The Ohio State 
University, Pennsylvania State 
University, Shanghai 
Astronomical Observatory, United 
Kingdom Participation Group, 
Universidad Nacional Aut\'onoma 
de M\'exico, University of Arizona, 
University of Colorado Boulder, 
University of Oxford, University of 
Portsmouth, University of Utah, 
University of Virginia, University 
of Washington, University of 
Wisconsin, Vanderbilt University, 
and Yale University.
\par
This paper has used archival data
from the \textit{Herschel} mission. \textit{Herschel} is an ESA space observatory
with science instruments provided by European-led Principal Investigator consortia and with important participation from NASA.

\section*{Data availability}
The data underlying this article will be shared on reasonable request to the corresponding author.


\bibliographystyle{mnras}
\bibliography{main}
\end{document}